\documentclass[a4paper,twocolumn,11pt,accepted=2022-04-05]{quantumarticle}
\pdfoutput=1
\usepackage[utf8]{inputenc}
\usepackage[english]{babel}
\usepackage[T1]{fontenc}
\usepackage{amsmath}
\usepackage{amssymb}
\usepackage{hyperref}
\usepackage[numbers,sort&compress]{natbib}

\begin{document}

\title{Entanglement Spectroscopy and probing the Li-Haldane Conjecture in  Topological Quantum Matter}

\author{Torsten V. Zache}
\email{Torsten.Zache@uibk.ac.at}
\affil{Institute for Quantum Optics and Quantum Information of the Austrian Academy of Sciences, Innsbruck 6020, Austria}
\affil{Center for Quantum Physics, University of Innsbruck, Innsbruck 6020, Austria}

\author{Christian Kokail}
\affil{Institute for Quantum Optics and Quantum Information of the Austrian Academy of Sciences, Innsbruck 6020, Austria}
\affil{Center for Quantum Physics, University of Innsbruck, Innsbruck 6020, Austria}

\author{Bhuvanesh Sundar}

\affil{Institute for Quantum Optics and Quantum Information of the Austrian Academy of Sciences, Innsbruck 6020, Austria}
\affil{JILA, Department of Physics, University of Colorado, Boulder CO 80309, USA}

\author{Peter Zoller}
\affil{Institute for Quantum Optics and Quantum Information of the Austrian Academy of Sciences, Innsbruck 6020, Austria}
\affil{Center for Quantum Physics, University of Innsbruck, Innsbruck 6020, Austria}

\begin{abstract}
	Topological phases are characterized by their entanglement properties, which is manifest in a direct relation between entanglement spectra and edge states discovered by Li and Haldane. We propose to leverage the power of synthetic quantum systems for measuring entanglement via the Entanglement Hamiltonian to probe this relationship experimentally. This is made possible by exploiting the quasi-local structure of Entanglement Hamiltonians. The feasibility of this proposal is illustrated for two paradigmatic examples realizable with current technology, an integer quantum Hall state of non-interacting fermions on a 2D lattice and a symmetry protected topological state of interacting fermions on a 1D chain. Our results pave the road towards an experimental identification of topological order in strongly correlated quantum many-body systems.
\end{abstract}

\maketitle

\begin{figure}[ht!]
	\centering{
		\includegraphics[width=\columnwidth]{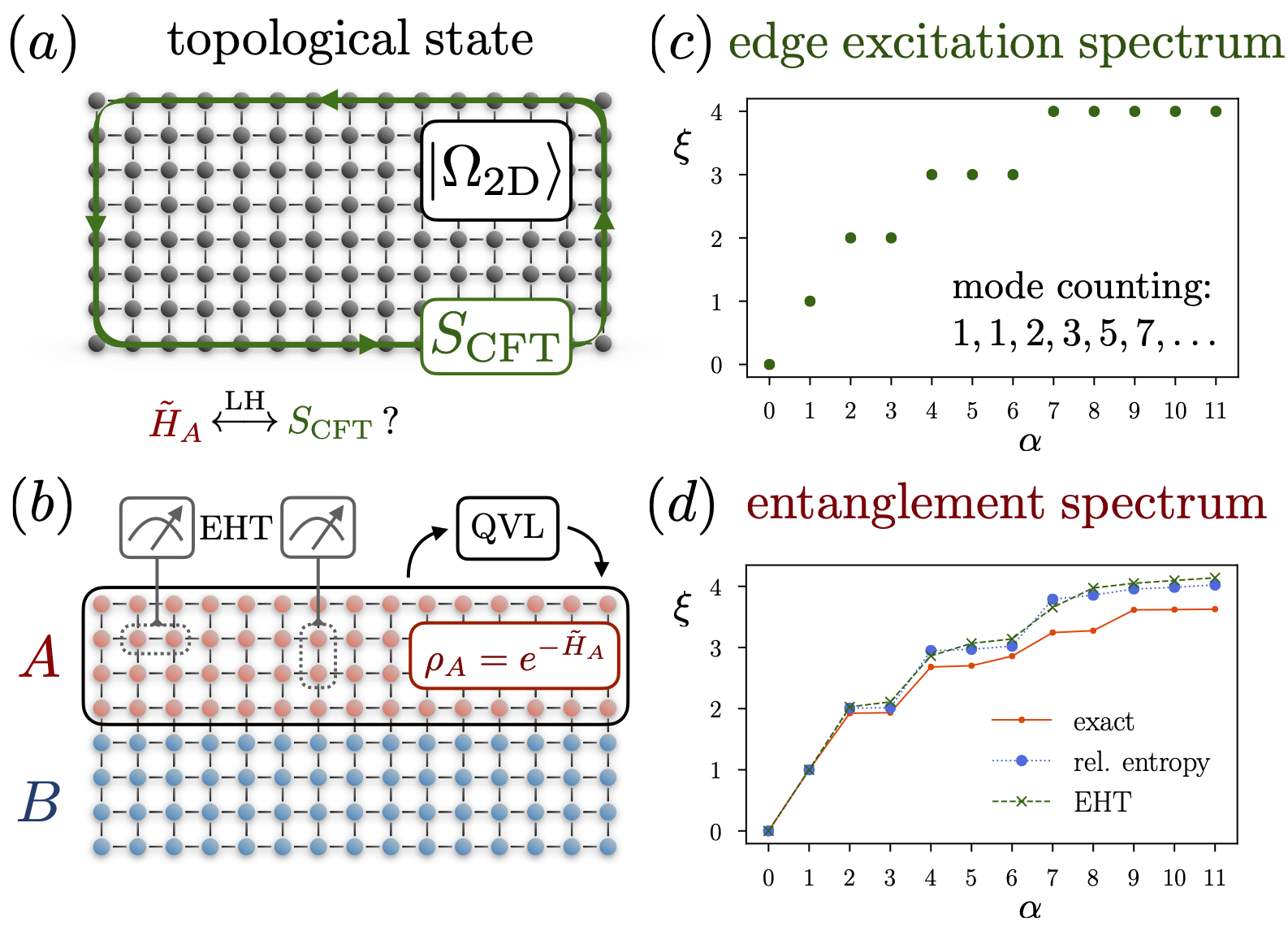}
		\caption{\label{fig:overview}
			{$(a)$ Consider a quantum Hall-like state $|\Omega_{\text{2D}}\rangle$, on a 2D lattice, that possesses chiral edge states described by a CFT with action $S_\text{CFT}$. $(b)$ We propose to perform entanglement spectroscopy based on Entanglement Hamiltonian Tomography (EHT) or Quantum Variational Learning (QVL) of the EH (see text) in order to probe the LH conjecture, which states that the reduced state on a subsystem $A(B)$ has an EH $\tilde{H}_{A(B)}$ whose low-energy spectrum is described by the same CFT as the edge states. $(c)$ The ideal CFT spectrum exhibits a linear dispersion with a characteristic counting of degenerate modes. $(d)$ In accordance with the LH conjecture, the low-lying levels of the ES obtained via EHT for a simulated experiment (green crosses), shown in comparison to the exact result (red dots) and an independent theoretical result based on an entropic principle (blue circles), exhibits the same structure as the CFT spectrum (see Sec.~\ref{sec:EHlearning} for more details).}
		}
	}
\end{figure}

\section{\label{sec:intro}Introduction}

Synthetic quantum matter provides us with a physical realization of topological phases~\cite{cooper2019topological,aidelsburger2018artificial,mancini2015observation,de2019observation,satzinger2021realizing,semeghini2021probing}. This can be implemented with ultracold atoms in optical lattices, realizing Bose or Fermi Hubbard models with artificial gauge fields~\cite{mancini2015observation,aidelsburger2015measuring,flaschner2016experimental,sompet2021realising}, and Rydberg tweezer arrays or trapped ions as spin models~\cite{de2019observation,scholl2021quantum,ebadi2021quantum}.
Since topological phases
cannot be detected by measuring a local order parameter~\cite{kitaev2006topological,levin2006detecting}, a key challenge is to identify  signatures of topological order (TO), which can be accessed in present experiments. As noted by Li and Haldane in context of the fractional Quantum Hall effect~\cite{li2008entanglement}, unique fingerprints of TO can be found in the entanglement structure of quantum states associated with topological phases. In the present paper we outline an experimentally realistic toolset of characterizing TO via measurement of entanglement properties. Our approach combines recently developed techniques for Entanglement Hamiltonian learning~\cite{kokail2020entanglement,kokail2021quantum}, and the Bisognano-Wichmann prediction~\cite{bisognano1975duality,bisognano1976duality} of the local structure of entanglement Hamiltonian into an efficient and universally applicable measurement protocol. In particular, our discussion will focus on `probing the Li-Haldane conjecture', i.e.~the observation of correspondence between the entanglement spectrum and edge state excitations as a signature of TO in 2D quantum Hall physics~\cite{regnault2017entanglement}.

The so-called Li-Haldane (LH) conjecture~\cite{li2008entanglement,chandran2011bulk,regnault2017entanglement} states that for a topological ground state $|\Omega \rangle$ -- describing, e.g., a quantum Hall system -- the low-lying levels of the Entanglement Spectrum (ES) are in one-to-one correspondence to the spectrum of the conformal field theory (CFT) that describes gapless excitations at the edge of the system~\cite{li2008entanglement,qi2012general} (see Fig.~\ref{fig:overview}). Here, the ES denotes the eigenvalues $\{\xi_\alpha\}$ of the Entanglement Hamiltonian (EH) $\tilde{H}_A$, which determines the reduced state
\begin{align}\label{eq:reduced_state}
	 \rho_A = \text{Tr}_B \left[|\Omega \rangle \langle \Omega |\right] = \sum_\alpha e^{-\xi_\alpha} |\Phi^A_\alpha \rangle \langle \Phi^A_\alpha  | = e^{-\tilde{H}_A}
\end{align}
for a spatial bi-partition $A:B$. TO is thus revealed in measurements of the ES.

Insight into the structure of the EH, and thus the ES, is provided by the Bisognano-Wichmann (BW) theorem~\cite{bisognano1975duality,bisognano1976duality}. For the vacuum $|\Omega\rangle$ of a (relativistic) quantum field theory, it identifies the EH 
\begin{align}
    \tilde{H}_A = \int d^dx \, \beta_A(x) \mathcal{H}(x) + \text{const}
\end{align}
as a {\em local deformation} of the system Hamiltonian \mbox{$H =\int d^dx \, \mathcal{H}(x)$}.
Here, $\beta_A(x)$ may be interpreted as a local inverse temperature, typically increasing linearly with the distance from the entanglement cut that defines the bi-partition~\cite{casini2011towards,cardy2016entanglement}. For a Hamiltonian $H$ with edge states, one can invoke the BW result to argue that the EH supports the same type of edge excitations at the entanglement cut, thereby providing a field-theoretic explanation of the LH conjecture~\cite{swingle2012geometric}. 

The BW prediction also holds with remarkably accuracy for ground states of discrete lattice models~\cite{dalmonte2018quantum,toldin2018entanglement,giudici2018entanglement,zhang2020lattice,eisler2020entanglement,zhu2019reconstructing}, described by a Hamiltonian $H = \sum_j h_j$ with quasi-local operators $h_j$, as realized in present quantum simulation experiments.
Thus, BW suggests an {\em efficient parametrization} of the corresponding reduced states as
\begin{align}
\rho_A(\boldsymbol{g}) &= e^{-\tilde{H}_A^\text{def.}(\boldsymbol{g})} \;, \\
 \tilde{H}_A^\text{def.}(\boldsymbol{g}) &= \sum_{j\in A} g_j h_j  + \text{const}\; \nonumber,
\end{align}
where the EH takes the form of a local deformation of the system Hamiltonian, parametrized by a small (polynomial in the size of $A$) number of parameters $\boldsymbol{g} = \{g_j\}$. This can be exploited for an efficient reconstruction of $\rho_A$ via learning of the EH~\cite{anshu2021sample}, and in particular also for topological ground states of lattice models. For example, Entanglement Hamiltonian tomography (EHT)~\cite{kokail2020entanglement} finds the EH by fitting the ansatz to experimentally measured observables in the subsystem $A$. Alternatively, by realizing $\tilde{H}_A^\text{def.}(\boldsymbol{g})$ as an ansatz of the EH, quantum variational learning (QVL)~\cite{kokail2021quantum} finds an optimal approximation of the EH, enabling a subsequent measurement of the ES through spectroscopy. These protocols provide the missing link of the theoretical concepts (due to LH and BW) to experimental reality.

In this article, we demonstrate that by combining these ideas, it becomes possible to probe the LH conjecture experimentally, as illustrated in Fig.~\ref{fig:overview}. For simplicity, we first focus on an integer quantum Hall state of cold neutral fermions in an artificial magnetic field (Sec.~\ref{sec:LH_IQH_CA}). In this case the EH is quadratic in the fermions, such that we can efficiently analyze its 
properties using EHT. As a second example (Sec.~\ref{sec:spin_chain}), illustrating that our approach to probe the LH conjecture remains applicable for interacting systems, we study a symmetry protected topological (SPT) phase of a 1D chain where we employ QVL. In either case, we numerically confirm the validity of a BW-type structure of the EH and demonstrate the LH correspondence. We further discuss the experimental ingredients required to carry out our proposal and analyze its scalability. 
Finally in Sec.~\ref{sec:conclusion}, we conclude with a discussion of future perspectives, including a road-map towards an experimental identification of TO in strongly correlated quantum many-body systems, such as quantum spin liquids~\cite{szasz2020chiral,verresen2020prediction}. The appendices contain pedagogical derivations of the BW theorem and the LH conjecture, as well as details about our analysis of the presented models.

\section{\label{sec:LH_IQH_CA}LH conjecture for an integer quantum Hall state with cold atoms}
The entanglement structure of the ground state $|\Omega \rangle$ of a many-body Hamiltonian $H$ is imprinted on the EH $\tilde{H}_A$ according to Eq.~\eqref{eq:reduced_state}. In a general sense, the LH conjecture formulates the remarkable statement that the low-lying levels of the ES, i.e. the ``low-energy'' spectrum of $\tilde{H}_A$, stand in one-to-one correspondence with the physical low-energy spectrum of $H$ in the presence of a boundary.

In the following subsections, we provide a brief overview over this phenomenology and then discuss our proposal to probe the LH conjecture in detail. Guided by experimental capabilities, we focus on the fermionic Harper-Hofstadter model, described by the Hamiltonian
\begin{align}\label{eq:H2D}
	H_{\text{2D}} = &-\sum_{\mathbf{n}}  \sum_{j=x,y} t_{j} \left( e^{i\varphi^j_{\mathbf{n}}} c^\dagger_{\mathbf{n}} c_{\mathbf{n} + \mathbf{e}_j} + \text{h.c.}\right)  \nonumber \\ &-\sum_{\mathbf{n}}  \mu \, c^\dagger_{\mathbf{n}} c_{\mathbf{n}}  \;,
\end{align} 
where $c_{\mathbf{n}}^{(\dagger)}$ are fermion annihiliation (creation) operators on a $N_x \times N_y$ lattice with sites $\mathbf{n} = (n_x, n_y)$ and unit lattice vectors $\mathbf{e}_j$. Here, $t_j$ determine the tunneling amplitudes and we set $\varphi^x_{\mathbf{n}} =0$ and $\varphi^y_{\mathbf{n}} = 2\pi n_x \Phi$, 
corresponding to a magnetic flux $\Phi$ per unit plaquette. 
We emphasize that the following discussion of entanglement properties does not depend on the chosen gauge to represent this flux. In particular, the ES that we are interested in can be shown to be gauge-invariant, while the EH itself transforms appropriately (see App.~\ref{app:fermionic_gauss} for details).

Below, we first review
how to realize $H_{\text{2D}}$ in state-of-the-art quantum simulation experiments (Sec.~\ref{subsec:implementation}). In Sec.~\ref{subsec:LH_IQHE}, we discuss the edge states and the ES for this model, which serves as an introduction to the LH conjecture. Our main results are presented in Sec.~\ref{subsec:IQH_EHT}, where we demonstrate how to measure the EH and the ES using EHT, made efficient due to the free-fermion nature of the model. Comparing to the CFT of the known edge states, we quantify the required experimental resources to test the LH conjecture.

\begin{figure*}
	\centering{
		\includegraphics[width=2.09\columnwidth]{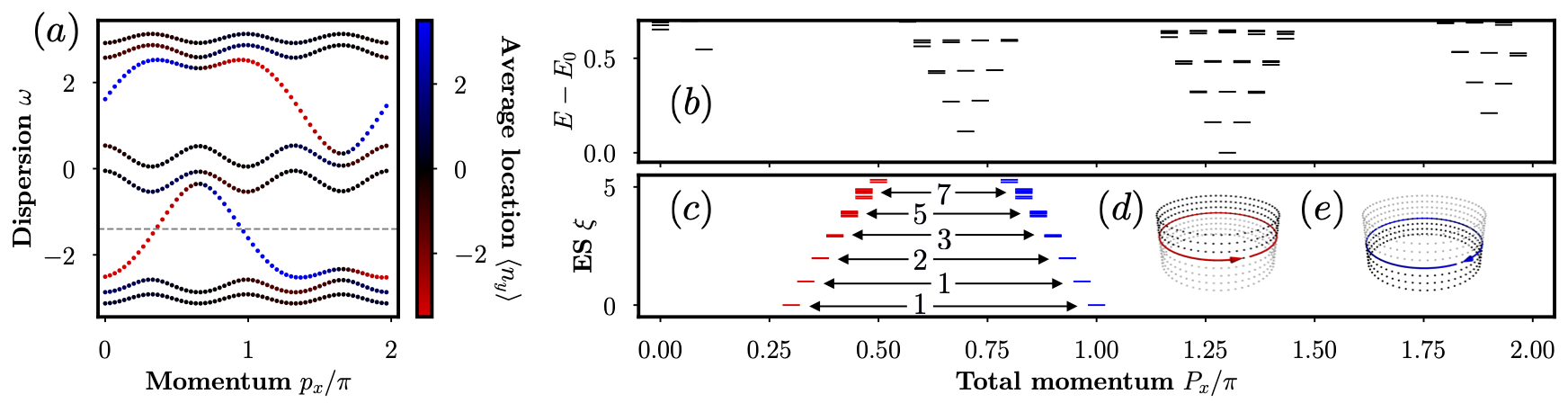}
		\caption{\label{fig:LH_}
			$(a)$ Single-particle dispersion relation for a cylinder of size $N_x \times N_y = 60 \times 8$ with $(t_y, \Phi) = (1.34,{0.335})$. All energies are measured in units of $t_x=1$, the lattice momentum is measured in units of the lattice spacing, i.e. $p_x \in [0,2\pi]$. The grey dashed line indicates the Fermi energy $E_F = -1.4$ corresponding to the filling $n_F = 0.3375$. The red (blue) colors show the average location of the single-particle states towards the lower (upper) end of the cylinder.
			$(b)$ Many-body excitation spectrum, for fixed total particle number $N_F$, above the ground state determined by filling all single-particle levels up to $E_F$ in $(a)$, plotted against the total momentum $P_x\in [0,2\pi]$.
			$(c)$ Entanglement spectrum obtained by tracing out the upper (red) and lower (blue) half of the cylinder in the sector with $N_F/2$ particles. The spectra are presented as ``universal ratios''~\footnote{Unless stated otherwise, all ES shown in this article are presented in this way.}, i.e. they are rescaled such that the two lowest lying levels are $\xi_0 = 0$ and $\xi_1 = 1$, and we indicated the number of levels per approximately degenerate multiplet. The two partitions of the cylinder and the expected chiral edge states are illustrated in the insets $(d)$ and $(e)$.
		}
	}
\end{figure*}

\subsection{\label{subsec:implementation}Cold atom implementation}
Systems of non-interacting fermions in an artificial magnetic field, as described by $H_\text{2D}$ [Eq.~\eqref{eq:H2D}], can be realized in a variety of different ways (see, e.g., \cite{aidelsburger2018artificial,cooper2019topological,goldman2016topological} for a review). Here we focus on realizations with cold atoms, in which case the Hamiltonian $H_\text{2D}$ can be implemented either in real-space, or using a synthetic dimension, which we briefly summarize below.

A ladder-like geometry with a long $x$-direction and a shorter $y$-direction
can be conveniently implemented using cold neutral fermions~\cite{celi2014synthetic,mancini2015observation} with several spin states (providing a ``synthetic'' $y$-direction) trapped in a one-dimensional optical lattice (the ``physical'' $x$-direction). 
While tunneling of the atoms along the physical direction sets the energy scale $t_x$
, the hopping in the synthetic direction can be achieved with Raman transitions induced by laser beams of wavelength $\lambda_R$ crossing at an opening angle $\theta$, which generates the desired phase \mbox{$\Phi =  a \cos (\theta)/\lambda_R$} with $a$ the lattice spacing of the optical potential
. For spin-$I$ atoms, the synthetic dimension has the maximal size $N_y=2I+1$ and the Raman transitions lead to an 
inhomogeneous tunneling strength \mbox{$t_{n_y,n_y+1} = (\Omega_R/2) \sqrt{I(I+1) - n_y (n_y+1)}$} with \mbox{$n_y = -I, \dots, I$} and Raman frequency $\Omega_R$.

While the synthetic dimension approach naturally favors ladder geometries, scaling up to a large homogeneous 2D system is more convenient with a real-space implementation. In this case, $H_\text{2D}$ can be realized by filling cold neutral atoms into a 2D optical potential, which is chosen to trap two different internal states in alternating columns of the lattice~\cite{jaksch2003creation}. Direct tunneling without changing the internal state is thus only possible along the columns, which sets $t_x$. Tunneling in the perpendicular direction $\propto t_y$ is again enabled by applying Raman lasers which drive transitions between the two internal states. Analogously to the synthetic dimension approach, this allows to imprint a phase on the hopping, which corresponds to the desired magnetic flux $\Phi$.

In either implementation, we assume the experimental capability to prepare a pure ground state, as well as to measure local few-body correlations. These requirements can be achieved, e.g., by adiabatic state preparation starting from a trivial product state, and by existing measurement techniques, such as spin- and site-resolved quantum gas microscopy~\cite{cooper2019topological,aidelsburger2018artificial,mancini2015observation,de2019observation,sompet2021realising,satzinger2021realizing,semeghini2021probing}.

Below, we consider both an idealized scenario of a cylinder geometry with a periodic $x$-direction, as well as open boundary conditions
which are naturally realized in cold atom experiments. 
For the latter case, we analyze the system sizes required to probe the LH conjecture in detail. In this context, we focus on an implementation with a necessarily short synthetic dimension, where we also take into account further experimental imperfections, such as a weak parabolic trapping potential in the physical direction or a weak linear Zeeman splitting of the states in the synthetic direction. Our results demonstrate that these imperfections constitute irrelevant perturbations that leave the physics of the LH conjecture unchanged.

\subsection{\label{subsec:LH_IQHE}Li-Haldane conjecture: edge states and entanglement spectrum}
To demonstrate the validity of the LH conjecture for the ground state $|\Omega_\text{2D}\rangle$ of $H_{\text{2D}}$, we need to calculate its ES and compare it to the spectrum of edge excitations in appropriate symmetry sectors. We start with the physical edge states for
chemical potential $\mu$ corresponding to a sector with fixed particle number $N_F = \sum_\mathbf{n} \langle c^\dagger_{\mathbf{n}} c_\mathbf{n}\rangle$ at filling \mbox{$n_F = N_F/(N_x N_y) \approx 1/3$} for $\Phi \approx 1/3$ and $t_y \approx 4/3 \times t_x$. 
While the bulk of $|\Omega_\text{2D}\rangle$ behaves as an insulator, the system develops gapless chiral edge modes in the presence of a boundary,
as evidenced by the dispersion relation on a cylinder shown in Fig.~\ref{fig:LH_}a.
The low-energy excitations at either edge are described by the effective CFT of a chiral boson~\cite{tong2016lectures}
\begin{align}\label{eq:CFT}
	S = \frac{m}{4\pi} \int dt \, dx \left[(\partial_t \phi) (\partial_x \phi) -v (\partial_x \phi)^2 \right] \;
\end{align}
for unit filling factor $\nu = 1/m = 1$. Here, $\phi$ is a compact field describing the right/left-moving excitations with velocity $\pm {|v|}$ determined by the slope of the dispersion at the Fermi surface. The many-body spectrum of $S$ consists of $1,1,2,3,5,7,11, \dots$ degenerate states with energies and momenta (relative to the ground state) given by \mbox{$E_k = {|v p_k|}$ and $p_k = \pm (2\pi k)/L$} for \mbox{$k = 0,1,2,3,4,5,6, \dots$}, where $L$ is the physical size of the periodic direction, i.e. the circumference of the cylinder (see App.~\ref{app:LH_derivation} for more details about the edge states and their mode counting). As illustrated in Fig.~\ref{fig:LH_}b, the many-body excitations on a finite lattice indeed reflect the structure expected from superimposing the two CFTs of both chiralities.

The LH conjecture states that the ES obtained by cutting the cylinder along the open direction into two halves reproduces the structure of the corresponding edge CFT. Fig.~\ref{fig:LH_}c shows the numerically obtained ES (see App.~\ref{app:fermionic_gauss} for a summary of properties of fermionic Gaussian states that underlie the numerics), in the sector with $N_F/2$ particles, sorted by momentum quantum numbers, which demonstrates the validity of the LH conjecture in the present scenario. Intuitively, the ground state contains information about the potential edge states at the virtual edge introduced by the entanglement cut. Tracing out one half of the system, the reduced state is dominated by the entanglement between a pair of ``virtual'' edge states~\footnote{In App.~\ref{app:density_response}, we discuss a quench experiment that makes these ``virtual'' edge states visible.} on either side of the cut, such that the CFT spectrum is imprinted on the low-lying ES (see Figs.~\ref{fig:LH_}d,e). For a rigorous presentation of this argument, we refer to \cite{qi2012general}.

For the interested reader, we give another self-contained explanation of the LH conjecture from a continuum field theory perspective in the appendix App.~\ref{app:LH_derivation}.
Our line of arguments can be summarized as follows (see also ~\cite{swingle2012geometric}): The universal properties of a topological state, such as relevant for the quantum Hall effect, are described by an effective topological field theory. By definition such a theory has a vanishing Hamiltonian density, except for an edge contribution on manifold with boundary. Applying the BW theorem, which states that the EH for a half-space partition of the ground state is given by a linear deformation of the system Hamiltonian, thus implies that the EH is dominated by the same edge contribution that also describes the edge states.
The BW theorem thus implies the LH conjecture.

\begin{figure}[t]
    \centering
    \includegraphics[width=\columnwidth]{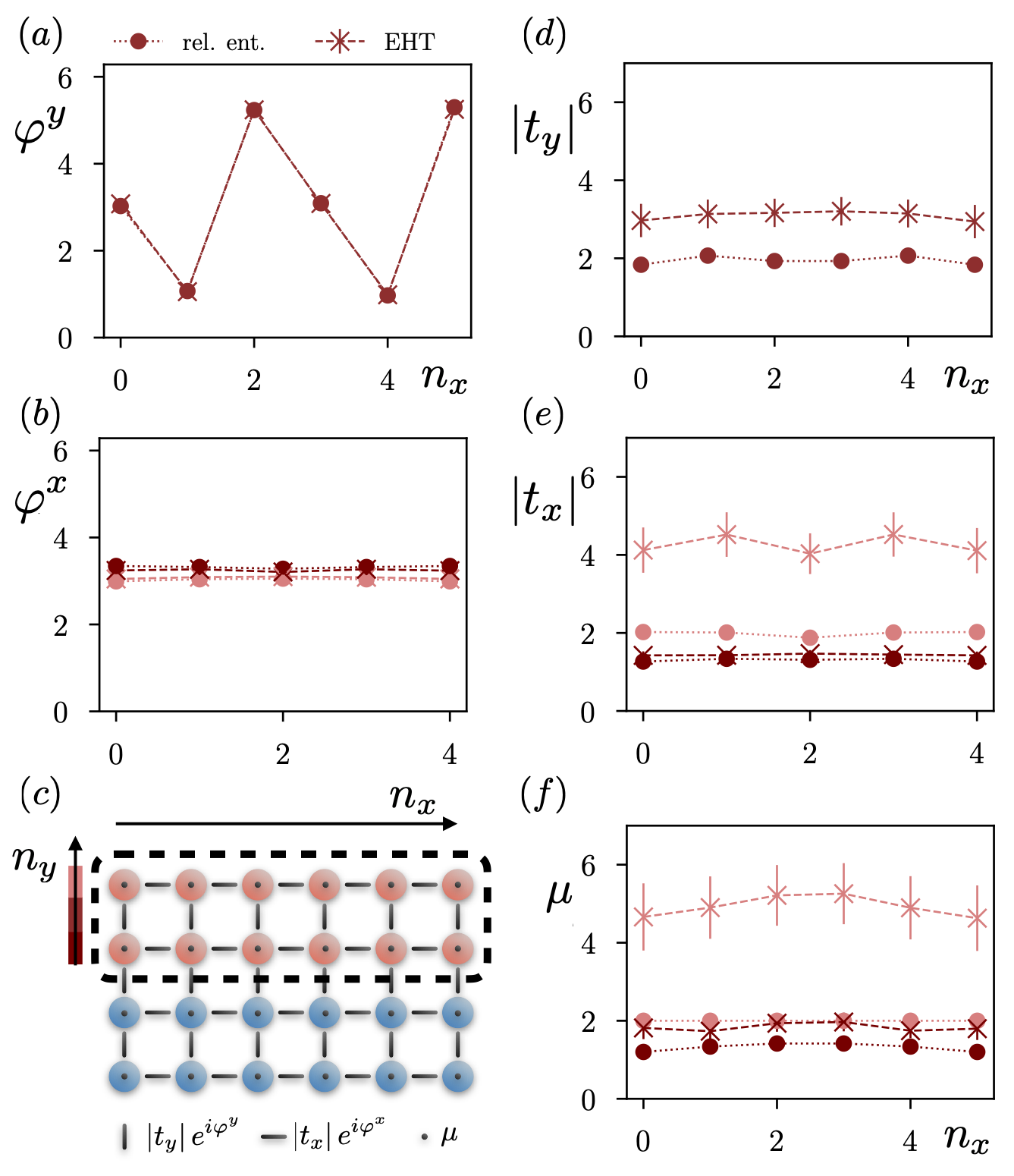}
    \caption{\label{fig:optimal_pars}Optimal parameters for the local BW-type ansatz of the EH [Eq.~\eqref{eq:local_ansatz}]. The different panels show the phase/amplitude of the tunneling terms in x- resp. y-direction [$(a/d)$ resp. $(b/e)$], as well as the local chemical potential terms [$(f)$] for a bipartition of the $N_x\times N_y=6\times 4$ lattice shown in $(c)$. We compare optimal parameters obtained from minimizing the relative entropy (circles) and EHT (crosses), see the labels in $(a)$ and $(b)$. The lightness of the symbols (dark to bright red) indicates the increasing distance $n_y$ w.r.t. the entanglement cut, as indicated in $(c)$. For EHT, the symbols correspond to the mean and its standard error obtained by repeating the simulation of the experiment $100$ times with a measurement budget of $N_\mathrm{meas.} = 10^4$.}
    \label{fig:optimal_parameters}
\end{figure}

\subsection{\label{subsec:IQH_EHT}Entanglement spectroscopy through EH learning via a Bisogano-Wichmann ansatz}
Instead of directly measuring the ES, our approach consists of first learning the EH and afterwards analyzing its spectral properties. This approach is motivated by the BW theorem which suggests to parametrize the EH as a local deformation of the system Hamiltonian (see App.~\ref{app:Bisognano-Wichmann} for a heuristic derivation of the BW result in the continuum). 

\subsubsection{Entanglement Hamiltonian Tomography}
In the following, we employ EHT~\cite{kokail2020entanglement} to learn the EH, which proceeds in two steps: $(i)$ choose a parametrized ansatz $\tilde{H}_A(\boldsymbol{g})$ for the EH and $(ii)$ determine parameters $\boldsymbol{g}$ by fitting the ansatz to observables $\mathcal{O}_A$ measured in the experimentally realized state $\rho_\text{exp.}$,
\begin{align}
     \text{Tr} \left[\mathcal{O}_A \, \rho_\text{exp.} \right]  &\equiv \left\langle \mathcal{O}_A \right\rangle_{\text{exp.}} \\
     &\overset{!}{=} \left\langle \mathcal{O}_A \right\rangle_{\boldsymbol{g}} \equiv \text{Tr}_A \left[\mathcal{O}_A \, e^{-\tilde{H}_A(\boldsymbol{g})} \right] \;. \nonumber
\end{align}
Step $(i)$ is efficient for a BW-type ansatz $\tilde{H}^{\text{def.}}_A(\boldsymbol{g})$ in the sense that the number of parameters scales only polynomially with system size (in contrast to the exponential scaling of a full state tomography). In full generality, step $(ii)$ is computationally hard because it requires to calculate expectation values for given parameters $\boldsymbol{g}$. However, depending on the structure of the ansatz, this step can be efficient as well.

Below, we apply EHT for the ground state $|\Omega_\text{2D}\rangle$ of Eq.~\eqref{eq:H2D}. Since Eq.~\eqref{eq:H2D} describes a system of non-interacting fermions, $|\Omega_\text{2D}\rangle$ and consequently all its reduced states $\rho_A$ are Gaussian, which implies that we can efficiently calculate correlation functions using Wick's theorem. In this context, also the second step of EHT is efficient. Simultaneously, the free-fermion structure enables an efficient diagonalization of the learned EH, thus providing access to the ES. For a generic interacting EH, such a classical diagonalization ceases to be feasible. We will come back to this case in Sec.~\ref{sec:spin_chain}.

Before turning to our results, let us briefly comment on the importance of chossing a local BW-type ansatz. Since we are dealing with Gaussian states, one can in principle calculate the EH exactly which takes the form \mbox{$\tilde{H}_A = \sum_{\mathbf{n}\mathbf{m}\in A} h^A_{\mathbf{n}\mathbf{m}} c^\dagger_\mathbf{n} c_\mathbf{m} + \text{const}$}. Here, the coefficients $h^A_{\mathbf{n}\mathbf{m}}$ are uniquely determined by the values of all two-point correlators $G_{\mathbf{n}\mathbf{m}} = \langle c^\dagger_\mathbf{n} c_\mathbf{m} \rangle$ in the subsystem (see App.~\ref{app:fermionic_gauss} for details), which suggests a direct approach to find the EH by measuring $G_{\mathbf{n}\mathbf{m}}$. However, this approach is not feasible in practice because the relation between $G_{\mathbf{n}\mathbf{m}}$ and $h^A_{\mathbf{n}\mathbf{m}}$ is numerically unstable against small (measurements) errors of $G_{\mathbf{n}\mathbf{m}}$ at large distances $|\mathbf{n}-\mathbf{m}|$, while measuring small values of $G_{\mathbf{n}\mathbf{m}}$ at large distances accurately is experimentally very challenging.
This problem is not present for our approach based on EHT because the assumed locality of the EH selects the most ``important'' contributions to the EH, as demonstrated by our results presented below.

Explicitly, we choose BW-type deformation
\begin{align}\label{eq:local_ansatz}
\tilde{H}_A^\text{def.}(\boldsymbol{g}) = &\sum_{\langle\mathbf{n} \mathbf{m}\rangle \in A} g^A_{\mathbf{n} \mathbf{m}} c^\dagger_\mathbf{n}c_\mathbf{m} \\ &+ \sum_{\mathbf{n}\in A} g^A_{\mathbf{n}} c^\dagger_\mathbf{n}c_\mathbf{n} + \text{const} \;, \nonumber
\end{align}
which includes only local ($\mathbf{n}\in A$) and nearest-neighbor ($\langle \mathbf{n}\mathbf{m}\rangle\in A$) terms,
as an ansatz for the EH. In view of experimental feasibility, we restrict ourselves to local measurements for determining the optimal parameters. Making use of the Gaussianity of the ansatz, we first calculate the reduced density operators for all pairs of neighboring lattice sites analytically and then fit the parameters to experimental probability distributions obtained from simulated projective measurements with a fixed number of repetitions $N_\text{meas.}$. For this procedure, we choose to measure in a particle number basis (and two rotated bases), which can be accessed experimentally by quantum gas microscopy (preceded by controlled single-particle tunneling). We refer to App.~\ref{app:EHT_QVL_summary} for more details.
For the results presented in the following, we focus on a realistic experimental setup with open boundary conditions, including non-homogeneous couplings, a trapping potential and a linear Zeeman shift.

\subsubsection{Results}
In Fig.~\ref{fig:optimal_parameters}, we illustrate the resulting structure of the learned EH. Here, we compare the EHT result to another set of optimal parameters found by a minimization of the relative entropy 
{$S(\rho_A|\tilde{\rho}_A)$ between the exact state $\rho_A$ and $\tilde{\rho}_A = e^{-\tilde{H}_A^\text{def.}(\boldsymbol{g})}$ determined by the ansatz} (see App.~\ref{app:fermionic_gauss} for details). For the entropic approach, we neglect shot noise due to a finite number of measurements.
{There are two main reasons for comparing our simulation of EHT to a minimization of the relative entropy: First, the consistency between the two variational solutions demonstrated below indicates robustness of the general approach against the detailed choice of cost function for the optimization. Second, the relative entropy (or its classical analog, the Kullback-Leibler divergence) is a common measure in quantum information theory and statistical inference, partly due to some useful mathematical properties (see also \cite{anshu2021sample}). In our case (without shot noise) this leads to a smooth and convex landscape for the parameter optimization that allows us to study scaling to larger system sizes more efficiently.}

The different panels of Fig.~\ref{fig:optimal_parameters} show the optimal EH parameters for a system of size $N_x\times N_y = 6\times 4$ for a bi-partition with subsystem size $N_x^A \times N_y^A = N_x \times N_y/2$  as illustrated in Fig.~\ref{fig:optimal_parameters}c. As seen from Fig.~\ref{fig:optimal_parameters}a and b, we find excellent agreement of the EHT and entropic results for the phases of the hopping terms. We find that the structure of these phases precisely reproduces the one of the original Hamiltonian [Eq.~\eqref{eq:H2D}]. For the remaining parameters (Fig.~\ref{fig:optimal_parameters}d-f), we find qualitative agreement of both approximations, with deviations increasing with the distance from the entanglement cut. Larger deviations at large distances are expected because the reduced state is more sensitive to changes in small parameters $\boldsymbol{g}$, while the EH parameters typically increase with increasing distance. Overall, the local approximations are consistent with expectations from the BW theorem, i.e. the EH is a deformation of the system Hamiltonian with parameters that are approximately constant parallel to the entanglement cut and exhibit an approximately linearly increasing slope in the perpendicular direction.

\begin{figure}[t]
	\centering
	\includegraphics[width=\columnwidth]{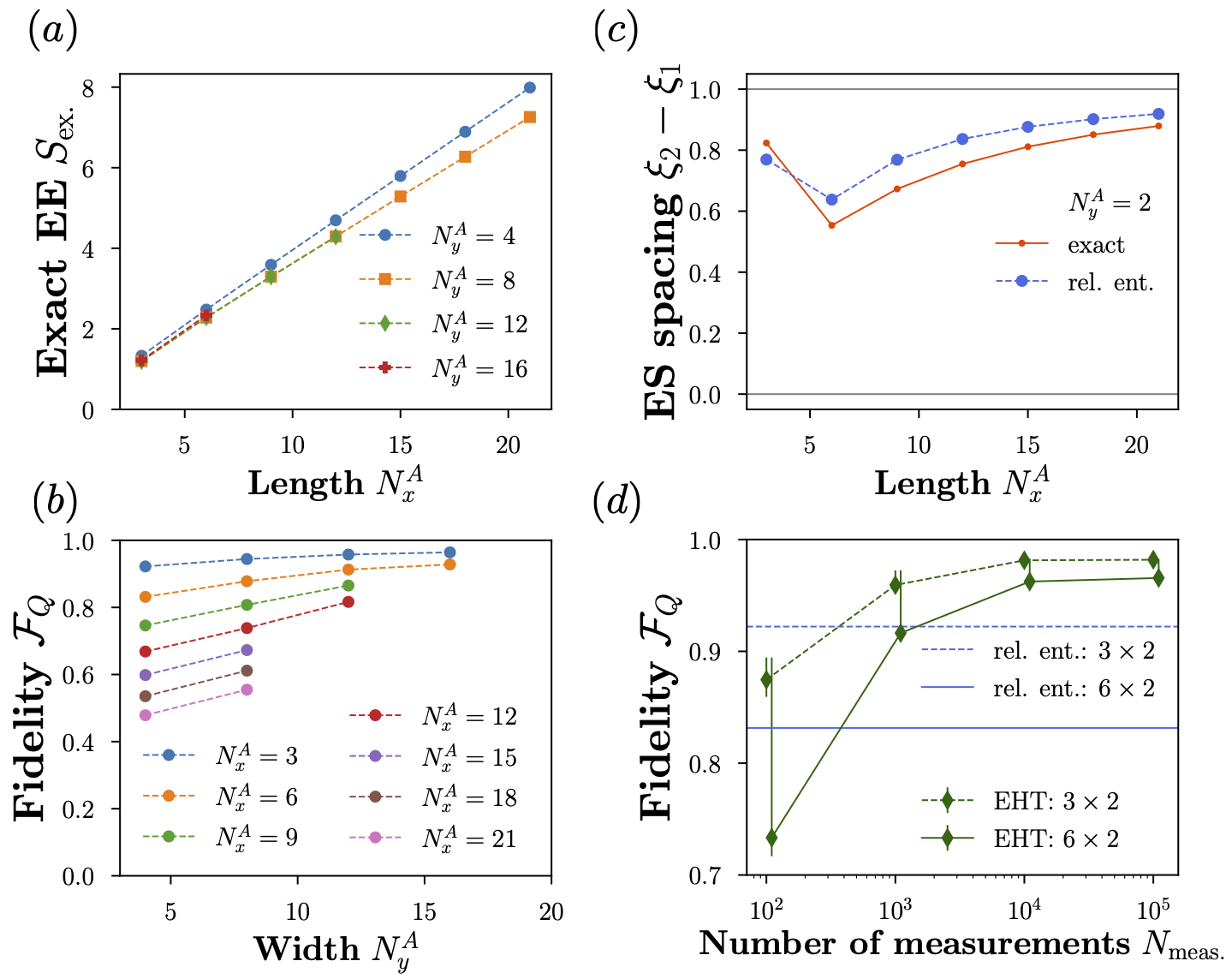}
 		\caption{\label{fig:IQH_local_approx} 
        $(a)$ The exact entanglement entropy exhibits linear scaling with $N^A_x$, approximately independent of $N^A_y$.
        $(b)$ The fidelity of the optimal approximation from minimizing the relative entropy slightly decreases with with $N^A_x$ while it increases with $N^A_y$, demonstrating the overall quality of the local approximation. 
        $(c)$ For increasing $N^A_x$, the spacing of consecutive ES levels approaches the degeneracy pattern predicted by the LH conjecture (shown is only the first non-trivial spacing which is expected to approach $1$).
        $(d)$ Scaling of the fidelity of the optimal approximation obtained from EHT as a function of the number of measurements $N_\mathrm{meas.}$ for different subsystem sizes $N^A_x\times N^A_y$. The data for $6\times 2$ corresponds to the same values of $N_\mathrm{meas.}$ as for $3\times 2$ and is just shifted for better visibility. The symbols correspond to the mean and its standard error obtained by repeating the simulation of the experiment $100$ times.
		}
\end{figure}

We proceed to study the quality of the approximate EH and analyze the scalability of our approach with system size and measurement budget {$N_\text{meas.}$}~\footnote{{To be precise, $N_\text{meas.}$ is the number of samples for each measurement basis in our simulations. For example, when measuring in the particle number basis,  $N_\text{meas.}$ corresponds to the number of snapshots obtained from, e.g., a quantum gas microscope.}} (Fig.~\ref{fig:IQH_local_approx}). Physically, we expect that the entanglement structure is dominated by virtual edge states at the entanglement cut. This is corroborated by the scaling of the exact entanglement entropy (EE), shown in Fig.~\ref{fig:IQH_local_approx}a, which is linear with the ladder length $N^A_x$ and approximately independent of the width $N^A_y$. 
Despite the fact that the reduced state is described by a small number of variational parameters  $N^\text{var} = 5 N^A_x N^A_y - 2(N^A_x + N^A_y)$, which scale only linearly with subsystem size, we find relatively large fidelities~\footnote{Here, we employ the non-logarithmic variety of the quantum Chernoff bound~\cite{audenaert2007discriminating,liang2019quantum} as a fidelity, which provides a natural distance measure among mixed quantum states.} for the entropic approximation (see Fig.~\ref{fig:IQH_local_approx}b). In general, the fidelity decreases mildly (sub-exponentially) with increasing size $N_x$ (parallel to the entanglement cut). Although the Hilbert space dimension is increasing exponentially, the fidelity even increases with increasing system size $N_y$ (orthogonal to the entanglement cut). As anticipated this behaviour is consistent with an EH dominated by states localized at the entanglement cut.

Even for small $N^A_y$, the low-lying ES corresponding to the local approximations, 
shown in Fig.~\ref{fig:overview}d for $N_x^A \times N^A_y = 21 \times 2$,
agree well with the exact one, which exhibits the onset of the degeneracy pattern predicted by the LH conjecture. Here, the EHT result is obtained for a single simulation of the experiment with $N_\mathrm{meas.}=10^4$ measurements. We find that both approximations give almost identical ES (in the shown particle number sector and for the low-lying levels), even though the fidelity of the EHT result ($\mathcal{F}_Q\approx 0.82$) is significantly higher compared to the entropic one ($\mathcal{F}_Q\approx 0.48$). 
As can be seen in Fig.~\ref{fig:overview}c), for a small system size the learned ES agrees better with expected structure of the CFT spectrum than the exact ES.
Therefore, in experiments where the exact ES is not known, a careful finite size analysis will be necessary in order to avoid premature conclusions (see also \cite{chandran2014universal}). Fig.~\ref{fig:IQH_local_approx}c shows the difference $\xi_2- \xi_1$, which according to the LH conjecture should approach $\xi_1-\xi_0 = 1$, as a function of the ladder length. This illustrates which system sizes will be required to resolve the characteristic structure of the low-lying ES, such as the equal spacing corresponding to a linear dispersion, for a given accuracy.

Since the scaling behaviour discussed above is consistent with physical expectations about the nature of the entanglement, we expect that it applies to both EHT and the entropic approximation. As can be seen in Fig.~\ref{fig:IQH_local_approx}{d} for relatively small system sizes, we find that for a sufficiently large measurement budget EHT provides a better approximation {(higher fidelity)} of the EH
than the entropic approach. {While this improved approximation might come as a surprise, we emphasize that in general there is no direct relation between the relative entropy and the fidelity. Since both quantities measure ``closeness'' of quantum states in a different way, an optimal variational solution w.r.t. the relative entropy can be sub-optimal w.r.t. the fidelity and vice versa. Note also that a direct optimization based on the fidelity is practically impossible due to the lack of an efficient protocol to measure the fidelity directly.}

In summary, our results indicate that a local approximation of the EH can be used to faithfully measure the entanglement structure for reasonably large system sizes and moderate number of measurements.
To probe the characteristic structure of the ES predicted by the LH conjecture, it suffices to realize a ladder geometry and analyze its entanglement for a bi-partition defined by cutting the ladder along its middle rungs. In general, the fidelity of the learned EH substantially increases with the width $N^A_y$ of the half-ladder, while the relevant low-lying levels of the ES converge with increasing length $N^A_x$. A reasonable compromise, corresponding to the final result presented in Fig.~\ref{fig:overview}d, is achieved for a subsystem size of $N_x^A \times N_y^A = 21 \times 2$ with a measurement budget of about $N_\text{meas.} = 10^4$. This puts probing the LH conjecture within reach of existing technology.

\section{\label{sec:spin_chain}Interacting chain of fermions or spins}
The results presented so far correspond to a system of non-interacting fermions, which we haven chosen for conceptual clarity. The simplicity of this example enabled the numerical benchmark simulations presented above, and makes an immediate experimental realization possible. However, as anticipated in the introduction, we are ultimately interested in strongly correlated many-body systems beyond the capabilities of classical computer simulations, where also EHT often ceases to be feasible. Specifically, the second step of EHT, the calculation of observables for given EH parameters, can become inefficient for large subsystems. But even if the learning of the EH is successful, obtaining its spectrum by classical diagonalization is in practice impossible.

To avoid these problems, we propose to use QVL~\cite{kokail2021quantum} as an alternative to learn the EH, which simultaneously provides an experimental realization of the EH. This approach has the advantage that also the structure of the ES can be obtained efficiently by performing physical spectroscopy.
In this section, we demonstrate that our proposal to probe the LH conjecture indeed remains applicable for interacting systems by relying on QVL instead of EHT. We illustrate this procedure for an example of a non-integrable lattice model in 1D which hosts an SPT phase. {The results presented in this section are directly relevant for recent experimental realizations of related models~\cite{de2019observation,sompet2021realising}.}

\begin{figure}[t]
	\centering{
		\includegraphics[width=\columnwidth]{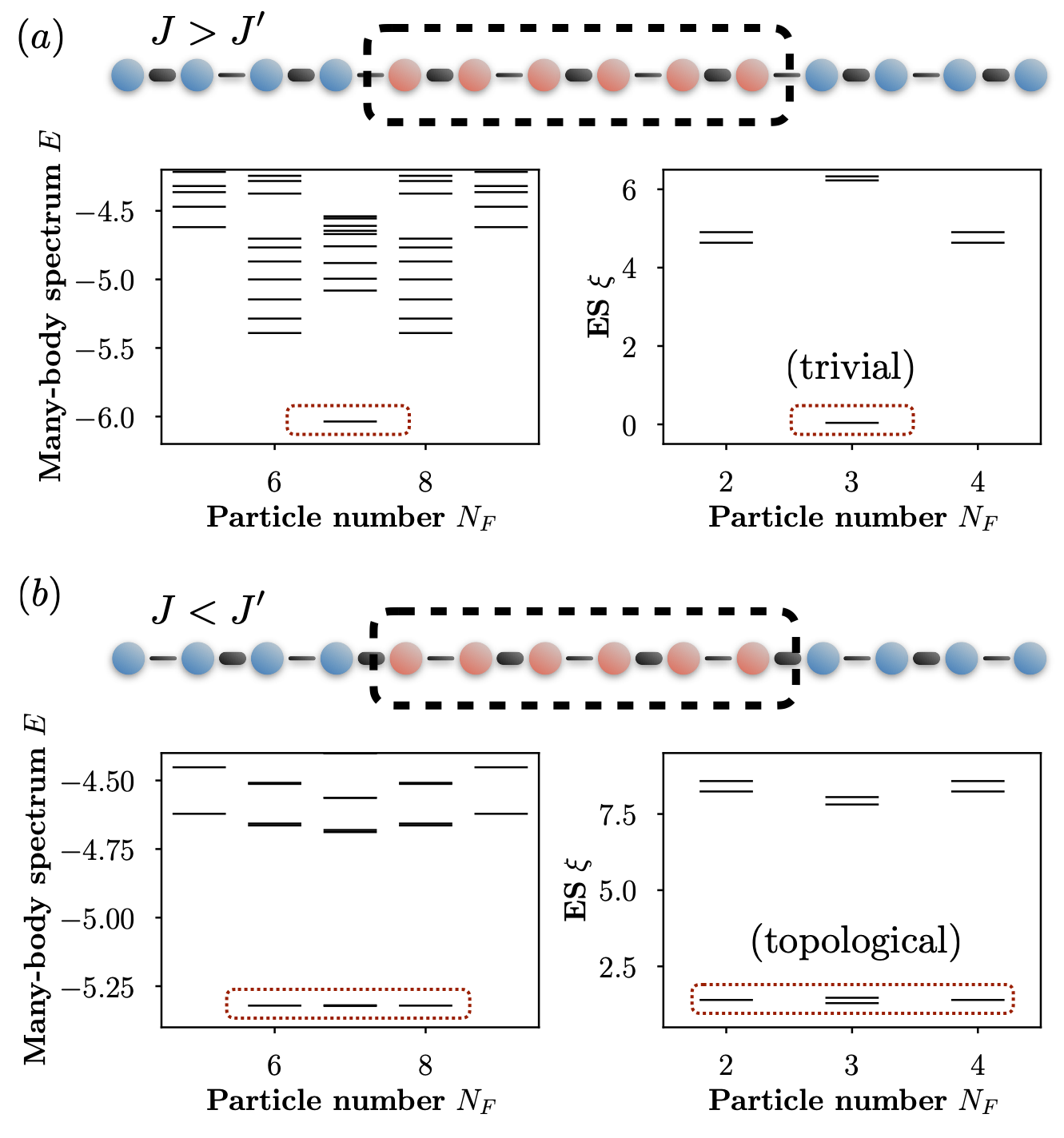}
		\caption{\label{fig:LH_SPT}
			Lowest lying levels of the many-body spectrum and entanglement spectrum for the trivial phase [$(a)\,, \;J>J'$] and the SPT phase [$(b)\,,\; J<J'$] for a finite chain of length $N=14$. The ES corresponds to the ground state with total particle number $N_F= 7$, tracing out the outer regions of the system, as illustrated above the plots. All data is obtained for $\delta = 0.5$ with $J/J' = 2$ and $1/2$ for $(a)$ and $(b)$, respectively. Here and in the following, all energies are measured in units of $J^2 + (J')^2 = 1$. The ground state manifolds of the ES show the same number of levels as the physical ground state manifolds, as indicated by the red dotted boxes, in accordance with the LH conjecture. Note that the ES is shown in absolute values, i.e. $\sum_\alpha e^{-\xi_\alpha} = 1$.
		}
	}
\end{figure}

\subsection{\label{sec:LH_SPT}LH conjecture for an SPT state}
We consider a 1D chain of interacting spin-less fermions described by
\begin{align}\label{eq:H1D}
    H_{\text{1D}} = &-\sum_n \left[J_n\left(c_n^\dagger c_n + \text{h.c.}\right)+ \mu \, c_n^\dagger c_n\right] \\
    &+\sum_n U_n \left(c_n^\dagger c_n - \frac{1}{2}\right)\left(c_{n+1}^\dagger c_{n+1} - \frac{1}{2}\right) \nonumber
\end{align}
for given chemical potential $\mu$ that fixes the total particle number $N_F = \sum_n c^\dagger_n c_n$.
In the following we specify bond-alternating tunneling elements $J_n = (J+J')/2 +(-1)^n (J-J')/2$ and interactions $U_n = 2\delta J_n$, and set $\mu=0$ (half-filling). The model is equivalent to the bond-alternating XXZ spin chain studied in~\cite{elben2020many},
\begin{align} \label{eq:H1D_spin}
		H_{\text{1D}} = \sum_{n} \left[\frac{J_n}{2} \left(\sum_{j=x,y}\sigma^j_{n} \sigma^j_{n+1} + \delta \, \sigma^z_{n} \sigma^z_{n+1} \right) \right]
\end{align}
with Pauli matrices $\sigma_n^{x/y/z}$ representing the spin-$1/2$ degrees of freedom.
For a vanishing anisotropy, $\delta=0$, the model reduces to the non-interacting SSH model, but for generic $\delta$ the model is interacting and non-integrable~\cite{elben2020many}. Its phase diagram for open boundary conditions has been obtained in~\cite{elben2020many} by measuring a many-body topological invariant, revealing a transition between a trivial and an SPT phase at $J=J'$, as well as a conventional antiferromagnetic phase for sufficiently large $\delta$.

Let us focus on the trivial and SPT phases.
Both phases are gapped, but while the ground state is unique in the trivial phase, the SPT phase develops a four-fold degeneracy in the thermodynamic limit. This is illustrated in Fig.~\ref{fig:LH_SPT}, where
the many-body spectrum, sorted by the total particle number $N_F$ (equivalently the total magnetization in the spin representation),
is shown for a finite chain.
The degeneracy arises from zero energy excitations at either end of the chain, protected by internal symmetries (e.g. time-reversal in the present case). These edge states affect not only the physical spectrum, but also the ES, as shown in Fig.~\ref{fig:LH_SPT} for a bi-partition of the chain into a bulk of size $N=6$ and its complement in a chain of length $N=14$. Indeed, the ``lowest-energy'' structure of the ES reproduces the (non-)degeneracy of the SPT (trivial) phase, which can be interpreted as a 1D instance of the LH conjecture. As a consequence, it is possible to detect this topological phase by measuring the ES.

\subsection{\label{sec:EHlearning}EH learning via QVL}
We now apply QVL to learn the EH and subsequently analyze its spectral properties to probe the LH conjecture.
Similar to EHT, QVL proceeds in two steps: $(i)$ we make a BW-type ansatz $H_A^\text{def.}(\boldsymbol{g})$ for the EH, and $(ii)$ we determine a set of optimal parameters $\boldsymbol{g}$. For a finite subsystem in the bulk $A$, we chose the ansatz
\begin{align}
\label{eq:chain_ansatz}
    H_A^\text{def.}(\boldsymbol{\beta}) =\sum_{(n,n+1)\in A} \beta_{n,n+1} h_{n,n+1} \;,
\end{align}
where $h_{n,n+1}$ abbreviates the operators appearing in Eq.~\eqref{eq:H1D} or~\eqref{eq:H1D_spin} such that $\sum_n h_{n,n+1} = H_\text{1D}$. In other words, we assume that the EH takes the form of a spatial deformation of the system Hamiltonian, which simplifies the problem to finding the local ``inverse temperatures'' $\beta_{n,n+1}$ at every link.

In contrast to EHT, the second step of QVL consists of a feeback loop between a classical optimization of the parameters $\boldsymbol{g}$ and quantum processing step where we evolve the reduced state with the ansatz and monitor observables $\mathcal{O}_A$ in the subsystem,
\begin{align}
\langle O_A \rangle_t = \text{Tr}_A \left[ \rho_A \, e^{i H^\text{def.}_A t} \, O_A\, e^{-i H^\text{def.}_A t} \right] \;.
\end{align}
As shown in \cite{kokail2021quantum}, demanding that $\langle O_A \rangle_t$ remains constant enables an efficient learning of the EH. Here, we chose all nearest-neighbors correlators $\langle \sigma^x_j \sigma^x_k\rangle_t$ in the subsystem to achieve this goal (see App.~\ref{app:EHT_QVL_summary} for details).

\begin{figure}[t]
	\centering{
		\includegraphics[width=\columnwidth]{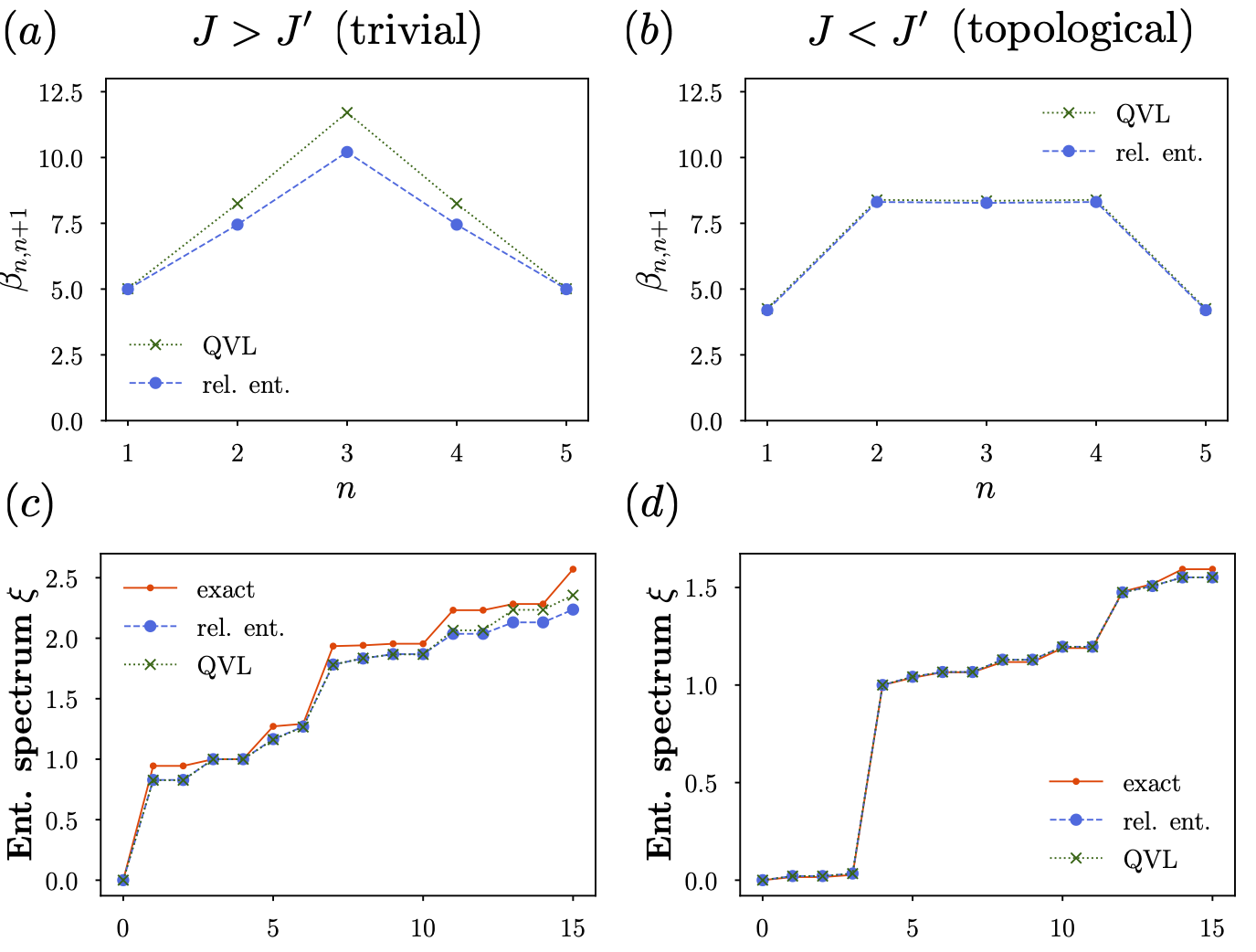}
		\caption{\label{fig:LH_SPT_variational}
		    The optimal EH parameters $\beta_{n,n+1}$ [see Eq.~\eqref{eq:chain_ansatz}] obtained from QVL are shown in $(a)$ and $(b)$ for the same setup presented in Fig.~\ref{fig:LH_SPT}, in comparison to an independent result obtained by minimizing the relative entropy. The corresponding reduced density operators for the QVL (entropic) results have a fidelity of $\mathcal{F} = 0.99882(0.99884)$ resp. $0.99981(0.99980)$ for $(a)$ resp. $(b)$ w.r.t. the exact reduced state. As a consequence, the corresponding ES (blue circles resp. green crosses), shown in $(c)$ and $(d)$, agree well with the exact ES (red dots). Note that the ES are shown as universal ratios, here we rescaled $\xi_4-\xi_0 = 1$ for an easier comparison between $(c)$ and $(d)$.
		}
	}
\end{figure}

The results of a numerical simulation of this approach to learn the EH, neglecting shot-noise~\footnote{{The effect of shot-noise for QVL has been studied in~\cite{kokail2021quantum}. Since we expect no fundamentally different behaviour for the present example, we neglect it here to simplify the numerical simulation.}}, are summarized in Fig.~\ref{fig:LH_SPT_variational}. We compare the QVL result to an independent estimate of the EH obtained by an entropic principle as in Sec.~\ref{sec:LH_IQH_CA}. Our findings demonstrate once again the validity of BW-type deformation of the EH. Despite the small subsystem sizes, we observe a qualitative difference between the trivial phase, where the deformation takes a triangular shape (Fig.~\ref{fig:LH_SPT_variational}a), and the topological phase, where the deformation exhibits a flat bulk (Fig.~\ref{fig:LH_SPT_variational}b). Since QVL is not sensitive to the overall rescaling of the ansatz, we have fixed this scale independently for this comparison (see App.~\ref{app:EHT_QVL_summary} for details). The triangular deformation can be interpreted as joining two linear slopes predicted by the BW theorem in the case of a large correlation length in the trivial gapped phase (see also \cite{eisler2020entanglement} for a study of the present model at $\delta=0$). Similiar to the non-interacting fermions discussed above, we find larger deviations for large values of $\beta_{n,n+1}$ because they play a less important role for reproducing the reduced state. For the SPT phase, we note that the shape of the deformation is very similar to the one observed recently in the topological phase of a $\mathbf{Z}_2$ lattice gauge theory in $2+1$ dimensions~\cite{mueller2021thermalization}.

The ES corresponding to the learned EHs are shown in Fig.~\ref{fig:LH_SPT_variational}c and d. We find excellent agreement with the exact ES, in accordance with the high fidelities of the learned EHs (see caption of Fig.~\ref{fig:LH_SPT_variational}). In particular, the (approximate) four-fold degeneracy in the SPT phase is clearly visible, while it is absent in the trivial phase, as predicted by the LH conjecture.

\begin{figure*}[t]
	\centering{
		\includegraphics[width=2.09\columnwidth]{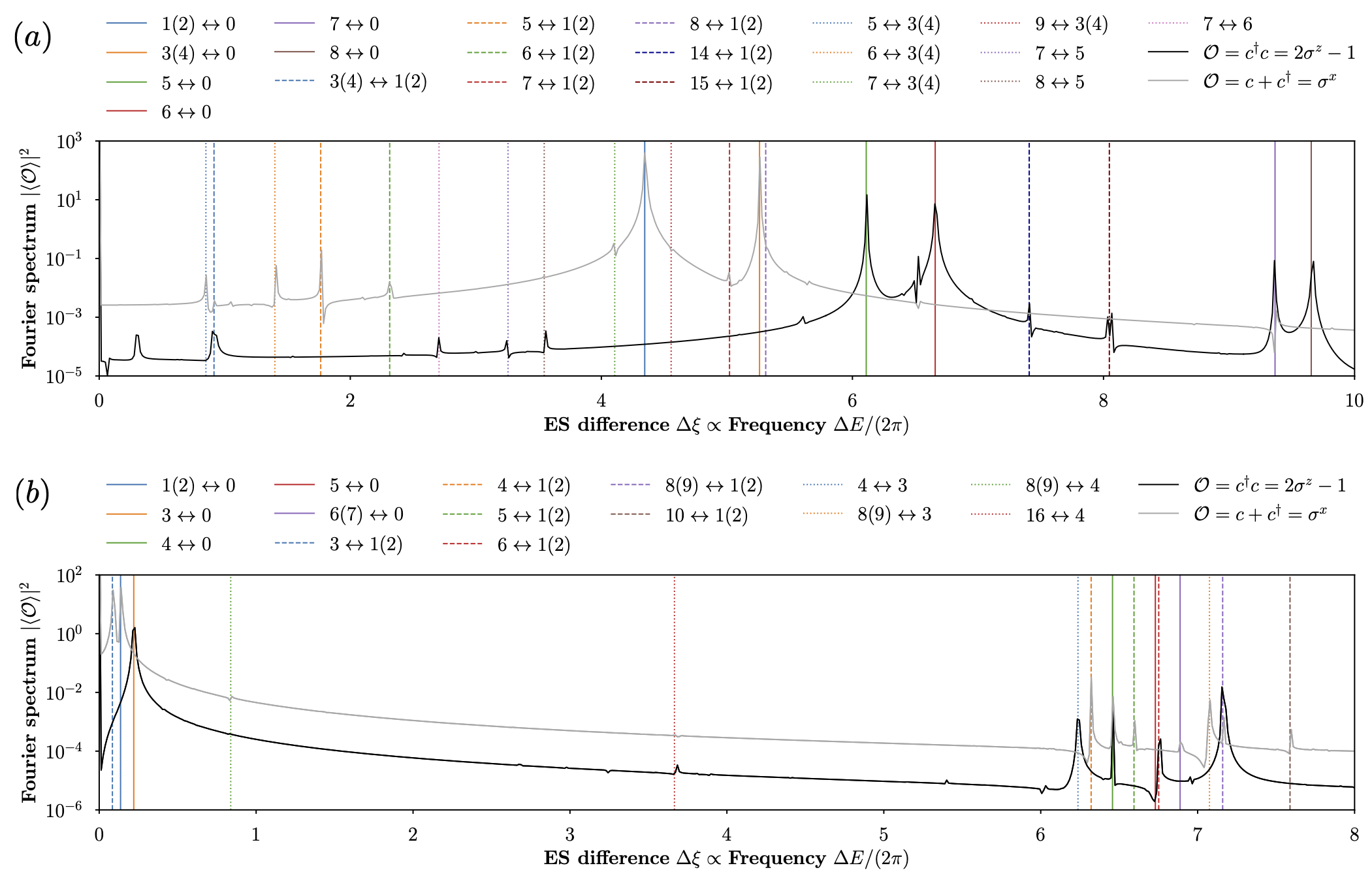}
		\caption{\label{fig:LH_SPT_spectra}
	    Spectroscopy of the trivial [$(a)$] and topological [$(b)$] states. The solid black (grey) lines show spectra obtained for an operator $\mathcal{O}$, indicated in the legend, time evolved with the learned EHs obtained via QVL (see Fig.~\ref{fig:LH_SPT_variational}), perturbed by the same operators (see text). The various colored vertical lines indicate expected transitions corresponding to two energy levels of the learned EH, as labelled in the legend. For example, the dominant resonance in $(a)$ for $\mathcal{O}=c^\dagger c$ around $\Delta \xi \approx 6$ corresponds to a transition between $\xi_0$ and $\xi_5$ (indicated by the solid green line), which are two lowest lying lying levels in the dominant particle number sector with $N_F = 3$ in the trivial phase (see Fig.~\ref{fig:LH_SPT}a).
		}
	}
\end{figure*}

\subsection{Entanglement Spectroscopy}
The ES shown in Fig.~\ref{fig:LH_SPT_variational}c/d have been obtained by classical diagonalization of the learned EH. As emphasized above, this approach becomes inefficient for large subsystems. In this section, we discuss the alternative of performing spectroscopy of the learned EH. To this end, we evolve the subsystem $\rho_A$ with the learned EH perturbed by another operator, $H = \tilde{H}^\text{def.}_A + \epsilon H'$, and measure the resulting dynamics of an observable $\langle\mathcal{O}\rangle(t)$. Analogously to ordinary atomic spectroscopy, the perturbation $H'$ induces transitions among different states in the ``thermal'' mixture $\rho_A = e^{-\tilde{H}_A}$. Resonances in the Fourier spectrum $|\langle\mathcal{O}\rangle(\omega)|^2$ thus correspond to ``energy'' differences in the ES.

The results of such a simulated spectroscopy are summarized in Fig.~\ref{fig:LH_SPT_spectra}. As usual, different spectroscopic protocols reveal different transition, which depend on the choice of observables and perturbations. Since the entanglement structure is dominated by edge contributions, we chose $H'$ and $\mathcal{O}$ close the entanglement cut for a strong signal. Here we have chosen $H' = \mathcal{O}$ for simplicity and set $\epsilon=0.1 (0.02)$ for the trivial (topological) state (see App.~\ref{app:EHT_QVL_summary} for details). As shown Fig.~\ref{fig:LH_SPT_spectra}, almost all low-lying levels of the ES are revealed by chosing $\mathcal{O}= c_j^\dagger c_j$ resp. $c_j+c_j^\dagger$ (equivalently $2\sigma_j^z - 1$ resp. $\sigma^x_j$ in the spin representation) with $j=4$ a site closest to the cut. Note that these choices correspond to transitions within a fixed particle number (magnetization) sector resp. transitions which change the sector by $\Delta N_F = \pm 1$. For a fermionic implementation, one typically has access to densities perturbations and measurements of local particle numbers, while an operator like $c_j+c_j^\dagger$ requires an auxiliary bath due to atomic particle number conservation. For a spin model implementation however, it is straightforward to realize the spectroscopy as discussed above.

{The results presented in Fig.~\ref{fig:LH_SPT_spectra} demonstrate that it is in principle possible to carry out entanglement spectroscopy from analyzing quench dynamics. We find this approach very appealing because it is universally applicable and in principle scalable to large system sizes. Nevertheless, we would like to mention some technical challenges and opportunities whose closer investigation we leave for future work. For instance, as seen in Fig.~\ref{fig:LH_SPT_spectra} the distinct shapes with varying weight/height of the different peaks stand in no obvious relation to the corresponding levels from the entanglement spectrum. These peak features in general depend on the absolute values of the ES, the chosen perturbation (its operator structure and the observation time), the precision of the learned EH, etc. It would be interesting to take these features into account to develop a more sophisticated analysis to extract the levels of the ES, rather than just looking at the peak positions. This becomes particularly relevant when no model is available to predict the ES with a few parameters. For a general ES, the situation is reminiscent to the spectroscopy of, say, an unknown atom. In practice, it will be necessary to probe the system using different perturbations with varying matrix elements between the different levels, eventually revealing more and more transitions until a complete picture emerges. Here, the fact the learned EH takes the form of a physical (quasi-local) Hamiltonian suggests that the same techniques employed in, e.g., atomic spectroscopy can help to analyze its spectrum.  }

To summarize, our results demonstrate the possibility to probe the LH conjecture via entanglement spectroscopy based on QVL. This approach remains applicable and efficient for generic interacting systems, given a BW-type deformation of the EH. Experimental requirements include the realization of spatially deformed system Hamiltonians, as well as access to appropriate perturbations and observables in order to perform the spectroscopy.

\section{\label{sec:conclusion}Outlook}
The relationship between entanglement and edge states as formulated by the LH conjecture is fundamental for our understanding of TO. 
The results that we have presented in this article demonstrate that state-of-the-art 
quantum simulators provide the unique opportunity to raise this theoretical concept to experimental reality. Our proposal combines the high degree of control over synthetic quantum to realize exotic phases of matter with a novel class of quantum information protocols, based on insight obtained from the BW theorem, to analyze these systems. 

The examples discussed above provide a clear road-map for future experiments: Starting from relatively simple free-fermion models where the basic phenomenology of the LH conjecture can be observed for exisiting setups, we expect that our ideas will find applications for increasingly complex situations, including SPT states and eventually (non)-abelian TO as in the fractional QHE.

Symmetries and the corresponding super-selection rules play a crucial role for labelling the ES in the context of the LH conjecture. Whether and how our proposal remains applicable to characterize constrained systems with an extensive amount of conserved quantities, such as lattice gauge theories or fracton models~\cite{casini2014remarks,shirley2019universal}, is an exciting theoretical question that we leave for future work. In this context, we note that the role of gauge-invariance is very simple for the example of free fermions in an artificial magnetic field that we considered above. In this case, the ES is independent of the choice of gauge to represent the background flux, while the operator content of the EH transforms accordingly such that a BW-type deformation emerges in any gauge.

\begin{acknowledgments}
We thank Rick van Bijnen and Mikhail Baranov for valuable discussions. Work in Innsbruck has been supported by the Simons Collaboration on Ultra-Quantum Matter, which is a grant from the Simons Foundation (651440, P.Z.), the European Union's Horizon 2020 research and innovation programme under Grant Agreement No.\ 817482 (Pasquans), and by LASCEM by AFOSR No.\ 64896-PH-QC.
{The data supporting the findings of this study and the numerical codes used to produce this data are available from the corresponding author, T.V.Z., upon reasonable request.}
\end{acknowledgments}

\appendix

\section{BW theorem and LH conjecture from a field theory perspective}
For completeness, we provide a self-contained, heuristic derivation of the Bisognano-Wichmann (BW) theorem and the Li-Haldane (LH) conjecture in this appendix. 
While the following discussion is not mathematically rigorous, it highlights the essential physical assumptions, and yields the correct results. We refer to \cite{witten2018aps} and references therein for a review on entanglement properties of quantum field theories.

\subsection{\label{app:Bisognano-Wichmann} The Bisognano-Wichmann theorem}
To be explicit, we consider the vacuum $|\Omega\rangle$ of a relativistic scalar field theory, i.e. $|\Omega\rangle$ is the ground state of the Hamiltonian
\begin{align}
    H &= \int_{\mathbf{R}^d} d^dx \, h_x \;, \\ h_x &= \frac{1}{2}\pi_x^2 + \frac{1}{2}\left(\nabla_x\varphi_x\right)^2 + V(\varphi_x) \nonumber
\end{align}
with $\left[\varphi_x,\pi_y\right] = i \delta^{(d)}(x-y)$ and potential terms $V$. The BW theorem states that the EH $\tilde{H}_A$ corresponding to the half-space $A\subset \mathbf{R}^d$ defined by $x_1\ge0$ is given by the linearly deformed Hamiltonian
\begin{align}
\tilde{H}_A = \int_{A} d^dx \, \beta_x h_x \;, && \beta_x = 2\pi x_1 \;.
\end{align}
We will show this by an equivalence of functional integral expressions for the reduced density operator
\begin{align}\label{eq:rhoA}
    \text{Tr}_B |\Omega \rangle \langle \Omega | = \rho_A = \frac{1}{Z_A} e^{-\tilde{H}_A} \;,
\end{align}
where $B\subset \mathbf{R}^d$ is the complement of $A$ and $Z_A$ is a normalization constant.

We start with the LHS of Eq.~\eqref{eq:rhoA}. In a basis of eigenstates of $\varphi$, i.e. $|\varphi\rangle = |\varphi_A, \varphi_B \rangle$, the matrix elements of $\rho_A$ can be written as
\begin{align}
    \langle \varphi^+_A | \rho_A |\varphi_A^- \rangle &= \int D\varphi_B \langle \varphi^+_A, \varphi_B | \Omega \rangle \langle \Omega |\varphi_A^-, \varphi_B \rangle \\
    &\sim \int_{\varphi^-_A}^{\varphi^+_A}  \mathcal{D} \varphi \, e^{-\int_{-\infty}^\infty d\tau \int_{\mathbf{R}^d} d^dx \, \mathcal{L}} \;,
\end{align}
where we dropped a normalization constant. Here, we introduced a representation of the vacuum wave functional as $\langle \varphi^+ |\Omega \rangle \sim \int^{\varphi^+} \mathcal{D}\varphi \exp \left( -\int_{-\infty}^0 d\tau \int_{\mathbf{R}^d} d^dx \, \mathcal{L} \right)$ with Lagrangian
\begin{align}
    \mathcal{L} = \frac{1}{2} (\partial_\tau \varphi)^2 + \frac{1}{2} (\nabla_x \varphi)^2 + V(\varphi) \;.
\end{align}
The matrix elements $\langle \varphi^+_A | \rho_A |\varphi_A^- \rangle$ are thus encoded in the boundary conditions of the functional integral, which specify the field values $\varphi^{+/-}_A$ shifted infinitesimally below/above in Euclidean time $\tau$ from the slice $A\subset \mathbf{R}^{d+1}$.

The key ingredient underlying the BW theorem is the Lorentz invariance of the vacuum, which here allows us to change coordinates under the integral as
\begin{align}
    \tau = r \sin(\theta) \;, && x_1 = r\cos (\theta) \;, 
\end{align}
with $r\in [0,\infty)$ and $\theta \in [0,2\pi]$, keeping $\bar{x} = {x_2, \dots, x_d}$ unchanged. Assuming that the measure $\mathcal{D}\varphi$ remains invariant and that the local form of the potential $V$ remains unchanged, we find
\begin{align}
    \langle \varphi^+_A | \rho_A |\varphi_A^- \rangle \sim \int_{\varphi^-_A}^{\varphi^+_A}  \mathcal{D} \varphi \, e^{ - \int_{0}^{2\pi} d\theta \int_0^{\infty} dr \int_{\mathbf{R}^{d-1}} d^{d-1}\bar{x}  \, \mathcal{L}'}
\end{align}
with the transformed Lagrangian
\begin{align}
    \mathcal{L}' =   \frac{1}{2r} (\partial_\theta \varphi)^2 + r \left[ \frac{1}{2} (\partial_r\varphi)^2 + \frac{1}{2} (\nabla_{\bar{x}}\varphi)^2 + V(\varphi)\right] \;.
\end{align}

Note that the term $\frac{1}{2r} (\partial_\theta \varphi)^2$ can also arise from a contribution of the form $\frac{r}{2} \pi^2 + i \pi (\partial_\theta \varphi)$ after performing a Gaussian integral over the conjugate field $\pi$. We can therefore interpret $\theta$ as a new time coordinate with periodicity $\beta = 2 \pi$ and $r$ as a spatial coordinate with range $r \ge 0$, such that the functional integral equivalently describes a thermal state at inverse temperature $\beta$ and Hamiltonian density $x_1 h_x$ on the region $A$. This proves Eq.~\eqref{eq:rhoA}.

We emphasize that since the above derivation requires Lorentz invariance, it's extension to other systems such as lattice models is not directly obvious. However, there are examples where a similar calculation based on corner transfer matrices provides a lattice analog of the BW theorem~\cite{peschel1999density,peschel1999density2,eisler2020entanglement}. For a general ground state of some finite lattice Hamiltonian, it may be possible to find a low-energy effective description in terms of a relativistic QFT, potentially perturbed by operators that break Lorentz invariance, such that the above calculations may serve as a leading-order starting point. If the corrections to the resulting EH (which are potentially non-local terms) are small, we expect that the EH can still be approximated by a quasi-local operator with a BW-type deformation. Numerical evidence, including the cases studied in this paper, strongly supports this picture for several examples~\cite{dalmonte2018quantum,toldin2018entanglement,giudici2018entanglement,zhang2020lattice,eisler2020entanglement,zhu2019reconstructing}.

\subsection{\label{app:LH_derivation}The Li-Haldane conjecture}
Assuming the validity of the BW theorem, we can also give a simplified derivation that explains the essence of the LH conjecture (see also \cite{swingle2012geometric,pretko2016entanglement} for related arguments). 
To be explicit, we consider $U(1)$ Chern-Simons theory at level $m$, described by the action
\begin{align}
    S_{\text{CS}} = \frac{m}{4\pi} \int d^3x \epsilon^{\mu \nu \rho} a_\mu \partial_\nu a_\rho \;,
\end{align}
which is a topological field theory that serves as an effective description for fractional quantum Hall states with $\nu = 1/m$ (or an integer quantum Hall state for $m=1$) in terms of emergent gauge fields $a_\mu$. For a space-time without a boundary, $ S_{\text{CS}}$ is invariant under $U(1)$ gauge transformations,
\begin{align}
    a_\mu \rightarrow a'_\mu = a_\mu + \partial_\mu \alpha \;.
\end{align}
For an introduction to Chern-Simons theories and their application to quantum Hall states we refer to the literature, e.g. \cite{tong2016lectures,dunne1999aspects}.

To apply the BW theorem, consider $S_{\text{CS}}$ on half of Minkowski space parametrized by $(t,x,y)$ with $y\le0$. From the variation of the action,
\begin{align}
    \delta S_\text{CS} &= \frac{m}{4\pi} \int d^3x \, \epsilon^{\mu\nu\rho} \left[\delta a_\mu \left(\partial_\nu a_\rho \right) - \left(\partial_\nu a_\mu\right)\right] \nonumber\\
    &+ \frac{m}{4\pi} \int dt \,dx \, \left.\left(a_x \delta a_t - a_t \delta a_x\right)\right|_{y=0} \;,
\end{align}
we see that the most general condition that eliminates the boundary term is 
\begin{align}
    a_t = v a_x \qquad \text{for} \qquad y=0   \;,
\end{align}
specified by a constant $v$. It follows that the equations of motion in the bulk are trivial, i.e. $f_{\mu\nu} \equiv \partial_\mu a_\nu - \partial_\nu a_\mu = 0$. Equivalently, the theory's Hamiltonian is zero in the bulk with the only non-vanishing contribution coming from edge excitations, $H_\text{CS} = \int dx \, dy \, \delta(y) \mathcal{H}_x$. To make this explicit, we choose a gauge where $a_t = va_x$ everywhere and treat $a_t$ as a Lagrange multiplier for the constraint $f_{xy} = 0$. In this gauge, the fields canonically conjugate to $a_{x/y}$ are $\pi_{x/y} = \pm \frac{m}{4\pi} a_{y/x}$ and we find
\begin{align}
    H_\text{CS} &= \frac{mv}{4\pi} \int_{y\le 0} dx \, dy \, \left[ a_x \partial_x a_y - a_y \partial_x a_x\right] \\
    &= \frac{mv}{4\pi} \int_{y\le 0} dx \, dy \, \partial_y \left[(\partial_x\phi)^2\right] \;,
\end{align}
where the last equality holds upon solving the constraint via $a_{x/y} = \partial_{x/y} \phi$. Performing the $y$-integral, only a boundary contribution remains, giving \mbox{$\mathcal{H}_x = \frac{mv}{4\pi} (\partial_x\phi)^2$}.

We can now apply the BW theorem for $A\subset \mathbf{R}^2$ defined by $y\le0$, which yields
\begin{align}
    \tilde{H}_A = 2\pi \int_A dx \,dy \, (-y) \, \delta(y) \mathcal{H}_x = \frac{2\pi^2}{\Lambda} \int dx \, \mathcal{H}_x \;,
\end{align}
where we formally regulated the ill-defined integral with a short-distance cutoff $\pi/\Lambda$. This demonstrates the strongest form of the LH conjecture: The EH is directly proportional to the Hamiltonian of the edge states, which precisely corresponds to the action given in Eq.~\eqref{eq:CFT} of the main text.

Above we have focused only on the universal aspects of the conjecture -- the exact form of the EH depends on microscopic details that define the reduced state $\rho_A$, encoded here in the constants $v$ and $\Lambda$. In general, for any real system, there will be further generic ``high-energy'' contributions to the EH~\cite{li2008entanglement} arising from corrections to the purely topological action assumed here.

\subsection{Edge mode counting}
Finally, we briefly discuss the counting of edge modes for the example introduced above. For a finite periodic boundary of length $L$, we write the action for $\phi_x = \sum_{k=-\infty}^\infty e^{ip_k x} \phi_k$ in momentum space with $p_k =2\pi k/L$ as
\begin{align}
    S = \frac{m}{2\pi} \int dt \sum_{k=1}^\infty \left[\left(\partial_t \phi_k\right) ip_k \phi_{-k} + v p_k^2 \phi_k \phi_{-k}\right] \;.
\end{align}
It follows that the edge theory consists of a set of harmonic oscillators with commutation relations
\begin{align}
    \left[\phi_k,\phi_{k'}\right] = -\frac{2\pi}{mp_k} \delta_{k,-k'} \;,
\end{align}
i.e. for $k>0$ we can interpret $\phi_k = \sqrt{\frac{mp_k}{2\pi}} b_k^\dagger$ as creation and $\phi_{-k} = \sqrt{\frac{m p_k}{2\pi}} b_k$ as annihilation operators. The Hamiltonian is given
\begin{align}
    H = \frac{mv}{2\pi} \sum_{k=1}^\infty p_k^2 \phi_k \phi_{-k} = v \sum_{k=1}^\infty p_k b^\dagger_k b_k
\end{align}
and its ground state (for $v>0$) is $|\Omega \rangle$ with $b_k|\Omega \rangle = 0$ for all $k>0$. The excited eigenstates $|\{n_k\}\rangle$ with energies $E(\{n_k\}) = \frac{mv}{2\pi} \sum_{k=1}^{\infty} n_k p_k$, corresponding to the occupation numbers $n_k$, follow as
\begin{align}
    |\{n_k\}\rangle \propto \prod_{k=1}^\infty \left(b_k^\dagger\right)^{n_k}|\Omega \rangle \;.
\end{align}
This proves the mode counting pattern $1,1,2,3,5, \dots$ with energies $E \propto 0,1,2,3,4, \dots$ as discussed in the main text.

\section{\label{app:fermionic_gauss}Fermionic Gaussian states}
Many of the results presented in the main text have been obtained using properties of fermionic Gaussian states, some of which we summarize in this appendix for reference. We restrict ourselves to particle-number conserving states $\rho$ determined by a $N \times N$ hermitian matrix $K_{jk} = K^*_{kj}$ as
\begin{align}
    \rho = \frac{1}{Z(K)} e^{-\sum_{jk} K_{jk}c^\dagger_j c_k} \;, 
\end{align}
where $\{c_j,c^\dagger_k\} = \delta_{jk}$. The constant 
\begin{align}
Z(K) = \text{Tr}  \left[ e^{-\sum_{jk} K_{jk}c^\dagger_j c_k}  \right] = \text{det} \left[ \mathbf{1} + T\right]
\end{align}
follows from the normalization $\text{Tr} \left[\rho \right] = 1$. Here, $\text{Tr}\left[\dots\right]$ denotes a trace over the $2^N$-dimensional Hilbert space of the operators $c_j$, while the determinant $\text{det} \left[\dots\right]$ acts on the $N$-dimensional space of the matrices $K$, $\mathbf{1}$ and the transfer matrix $T=e^{-K}$.

\subsection{Correlation functions and entanglement}
Gaussian states have the unique property that they are completely determined by their two-point correlations functions
\begin{align}
    G_{jk} = \text{Tr} \left[\rho \, c^\dagger_j c_k \right] \;,
\end{align}
i.e. higher-order correlation functions factorize according to Wick's theorem. Given a quadratic Hamiltonian $H = \sum_{jk} h_{jk}c^\dagger_j c_k$, with single-particle eigenstates $\phi^p_j$ and corresponding energies $\omega$, i.e. $\sum_k h_{jk}\phi^p_k = \omega_p \phi^p_j$, the ground state $|\Omega\rangle$ for a given Fermi-energy $E_F$ is a Gaussian state determined by the correlators
\begin{align}
G_{jk} = \sum_{p} \Theta(E_F-\omega_p) \left(\phi^{p}_j\right)^* \phi^p_k\;.
\end{align}
Since any reduced state of a Gaussian state is also Gaussian, we can calculate the corresponding EH exactly using the relation~\cite{peschel2003calculation}
\begin{align}
    K^T = \log \left[G \left(G-\mathbf{1}\right)^{-1}\right] 
\end{align}
among the matrix $G$ and the transposed matrix $K^T$, where the indices of $G$ have to be restricted to the desired subsystem. The corresponding many-body ES $\{\xi\}$ then follows as $\xi = \sum_{\tilde{p}} n_{\tilde{p}}  \epsilon_{\tilde{p}}$ by summing up the eigenvalues $\epsilon_{\tilde{p}}$ of $K$ for a given particle number configuration $n_{\tilde{p}} \in \{0,1\}$.

The direct relation between the EH/ES and the correlation functions also clarifies the role of gauge transformations of the form $c_j \rightarrow c_j e^{-i\varphi_j}$ as follows. Since the correlators transform as $G_{jk} \rightarrow e^{i(\varphi_j - \varphi_k)}$, we obtain
\begin{align}
\sum_k G_{jk} \phi^{\tilde{p}}_k &= \zeta_{\tilde{p}} \phi^{\tilde{p}}_j \nonumber \\ \rightarrow  e^{i\varphi_j}\sum_k G_{jk} \left(e^{-i\varphi_k} \phi^{\tilde{p}}_k\right) &= \zeta_{\tilde{p}} \phi^{\tilde{p}}_k \;.
\end{align}
This implies that the eigenvalues $\zeta_{\tilde{p}}$ of $G$ and thus also the eigenvalues $\epsilon_{\tilde{p}}$ of $K$ remain unchanged. Therefore the ES  $\{\xi\}$ is seen to be gauge-invariant, while the operators that contribute to the EH transform accordingly.

\subsection{Fidelity and relative entropy}
For two  mixed states $\rho$ and $\sigma$, a natural fidelity $\mathcal{F}_Q$, related to the quantum Chernoff bound~\cite{audenaert2007discriminating,liang2019quantum}, that quantifies their similarity is given by
\begin{align}
    \mathcal{F}_Q(\rho,\sigma) &= \min_{0 \le s \le 1}\text{Tr} \left[\rho^s \sigma^{1-s}\right] \\
    &= \min_{0 \le s \le 1} \det\left(\mathbf{1} + e^{-sK^\rho}e^{-(1-s)K^\sigma}\right)  \;,
\end{align}
where the last equality holds for two fermionic Gaussian states, determined by matrices $K^\rho$ and $K^\sigma$.

Another measure of statistical distinguishability is the relative entropy
\begin{align}
    S(\rho | \sigma) = \text{Tr} \left[ \rho \left(\log \rho - \log \sigma\right) \right] \;,
\end{align}
which for two fermionic Gaussian states takes the form
\begin{align}
    S(\rho | \sigma) = -S(\rho) + \sum_{jk} K^\sigma_{jk} G^\rho_{jk}  + \log Z(K^\sigma) \;,
\end{align}
where $S(\rho) =- \text{Tr} \left[ \rho \log \rho  \right]$ is the von Neumann entropy of $\rho$.

\subsection{Hamiltonian dynamics}
Finally, the quantum dynamics of fermionic Gaussian states under a an arbitrary quadratic Hamiltonian $H = \sum_{jk} h_{jk}c^\dagger_j c_k$ (including time-dependent coefficients $h_{jk}$) can be obtained by evolving the two-point functions according to
\begin{align}
    i\partial_t G_{jk} = \sum_l \left(G_{jl} h^*_{lk} - h^*_{jl} G_{lk}  \right) \;,
\end{align}
where $h_{jk}^*$ denotes the complex (not hermitian) conjugate of $h_{jk}$.

\section{\label{app:density_response}Observing virtual edge states}
In this appendix, we discuss a possibility to make the ``virtual'' edge states responsible for the structure of the ES visible.
Having determined the EH (e.g. from EHT), qualitative features of the low-energy ES can also be revealed in a quench experiment with the reduced state $\rho_A = e^{-\tilde{H}_A}$. Specifically, we propose to measure the density response \mbox{$\chi_A (x{,y},t) = i \text{Tr} \left\lbrace e^{-\tilde{H}_A} \left[ n(x,t), n({y},0) \right] \right\rbrace$} along positions $x$ at the entanglement edge, where $n(x,t)$ is the density operator in the Heisenberg picture with respect to the EH. Here, we switched to a continuum notation to make closer contact with the expected edge CFT. {Assuming translation-invariance, we set $y=0$ and focus on $\chi_A(x,t) = \chi_A(x,0,t)$ in the following.} If the LH conjectures holds true, we can apply the CFT [Eq.~\eqref{eq:CFT}] to describe the EH, leading to the prediction (see further below for its derivation)
\begin{align}\label{eq:CFT_response}
\chi_A(x,t) = i\int_{p,q>0} dp \, dq \; & e^{i(q-p)(x-vt)} \\ & \times\left[n_A(p) - n_A(q)\right] \; \nonumber
\end{align}
with $n_A(p) = \left(1+ e^{\beta v p}\right)^{-1}$, where the velocity $v$ and the constant $\beta$ depend on the units in which time evolution w.r.t. the EH is measured, such that $\beta v p = \xi_p$ are the single-particle levels of the ES.

The response function $\chi_A(x,t)$ is directly accessible in experiments by measuring the change of the densities, $\delta n(x,t)= \text{Tr} \left\lbrace e^{-\tilde{H}_A} \left[n(x,t) -  n(x,0)\right] \right\rbrace \propto \chi_A(x,t)$, after evolving the subsystem $A$ with the EH including a small perturbation of the local chemical potential at initial time $t=0$ and position $x=0$. By slight abuse of notation $n(x,t)$ now denotes the Heisenberg operator w.r.t. the perturbed EH. 

In the same experimental setup, one can perform an analogous response measurement at the physical edge of the full system yielding a response $\chi(x,t)$. While $\chi_A(x,t)$ contains the relevant information about the low-lying ES, the analogous quantity $\chi(x,t)$ reveals the physical edge states. Comparing the result from both measurements thus enables a direct test of the LH conjecture.

\begin{figure}[t]
	\centering
	\includegraphics[width=\columnwidth]{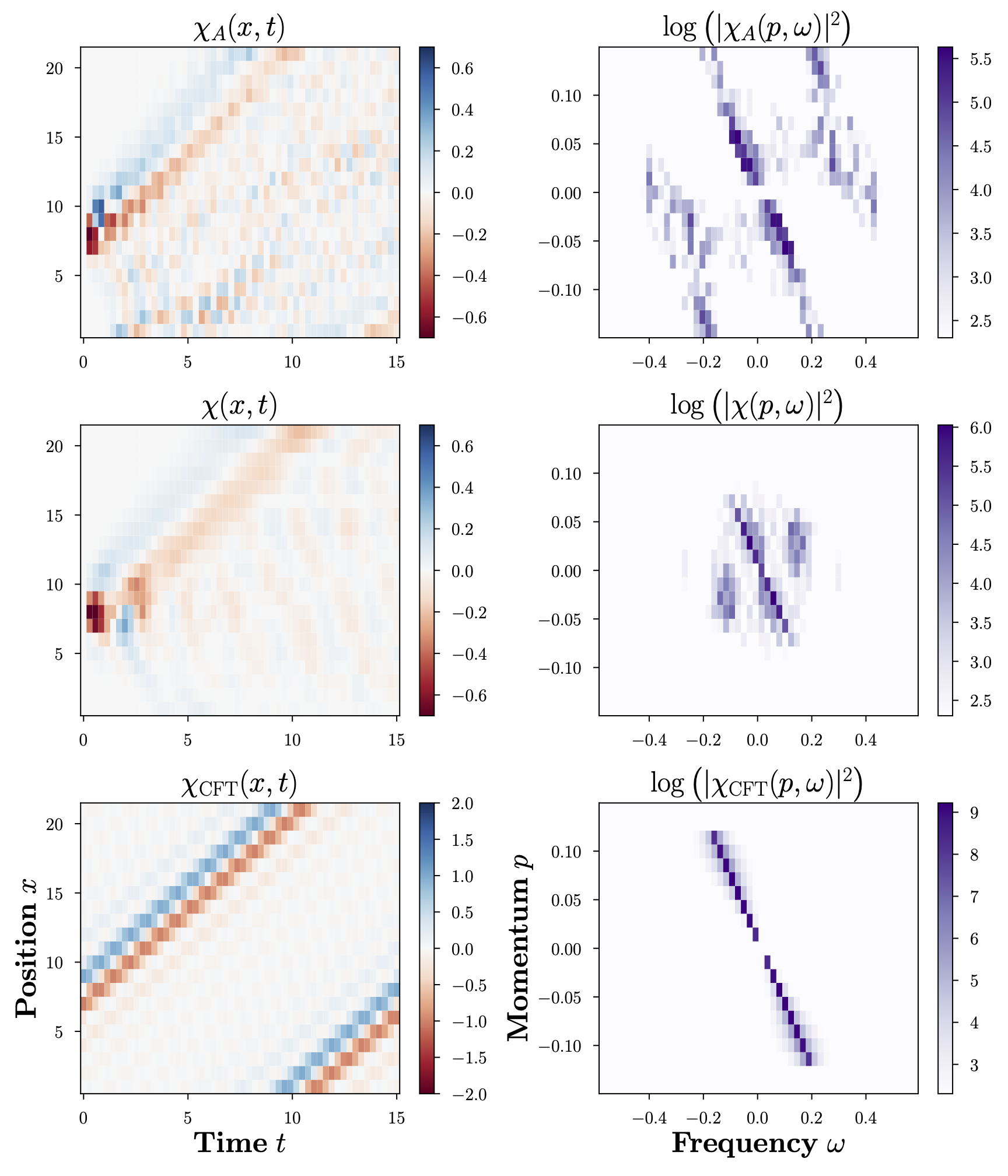}
	\caption{\label{fig:IQH_spectroscopy}
		Density response {$\chi$} in real space (left column) and its Fourier spectrum (right column). The (sub)system's response in the (top)middle row is in agreement with the CFT prediction (bottom). The spectra exhibit a clear ``chiral'' signal $\propto \delta(\omega - vp)$, while high-momentum/frequency contributions indicate corrections beyond the low-energy effective theory. For $\chi_A$, the time evolution is simulated with a local approximation of the EH obtained by minimizing the relative entropy. All simulations, corresponding to a $N_x \times N_y= 21\times 8$ lattice, are evaluated at $N_t=50$ time steps, and all response functions are normalized (in real space) to a maximum absolute value of one. Note that we have adjusted the color scales for better visibility. {Also note that the initial perturbation is placed at a position $y>0$ to reduce boundary effects.}
	}
\end{figure}

In Fig.~\ref{fig:IQH_spectroscopy}, we compare the results of a real-time simulation from which we extracted the response functions to the analytical expectation [Eq.~\eqref{eq:CFT_response}]. The real-space pictures clearly show a non-dispersing wave-packet travelling in one direction along the (entanglement) boundary, as expected for a chiral CFT. The anticipated linear dispersion $\omega(p) = vp$ is clearly visible in the Fourier spectra, and the corresponding structure of the low-lying states of the EH indeed coincides with those of the physical low-energy edges states.

\subsection{Derivation of the density response}
We now provide a derivation of Eq.~\eqref{eq:CFT_response}. To this end, we assume a reduced density matrix $\rho_A = e^{-\tilde{H}_A}$ with EH $\tilde{H}_A = \beta  \int^\infty_{0} dp\,   \omega(p) \, c^\dagger(p) c(p) + \text{const.}$ for inverse temperature $\beta$, linear dispersion $\omega(p) = vp$ and velocity $v$. The quench experiment proposed above corresponds to evolution under a time-dependent Hamiltonian $H_A(\lambda,t) = \tilde{H}_A / \beta + \lambda \delta(t) n(y)$ with $n(y) = c^\dagger(y)c(y)$ the particle number operator at a point $y$. In leading order time-dependent perturbation theory, the change $\delta n(x,t)$ of the density $n(x)$ at another point $x$ at a later time $t$ is given by 
\begin{align}
\delta n(x,t) &= \langle n(x,t)\rangle_\lambda - \langle n(x,t)\rangle_{\lambda = 0}\\
&= \lambda \chi_A(x,t) + \mathcal{O}(\lambda^2) \;,
\end{align}
where $\langle n(x,t)\rangle_\lambda$ denotes the expectation value of $n(x)$ under evolution with $H_A(\lambda,t)$ for the initial state $\rho_A$. Here, we introduced the linear response function
\begin{align}
    \chi_A(x,t) = i \text{Tr} \left\lbrace\rho_A \left[n(x,t),n(y,t)\right] \right\rbrace \;,
\end{align}
with the Heisenberg operators $n(x,t) = e^{i\tilde{H}_At/\beta} \, n(x) \, e^{-i\tilde{H}_At/\beta}$. Eq.~\eqref{eq:CFT_response} then follows from evaluating the commutator in Fourier space by applying Wick's theorem to the Gaussian state $\rho_A$, yielding
{
\begin{align}
    \text{Tr} &\left\lbrace\rho_A \left[c^\dagger(q) c(p),c^\dagger(r) c(l)\right] \right\rbrace  \nonumber \\
    &= \delta(p-r) \delta(q-l) \left[n_A(q) - n_A(p) \right] \;,
\end{align}}
 with ``thermal'' Fermi-Dirac distribution
\begin{align}
    n_A(p) = \left(1 + e^{\xi(p)}\right)^{-1}
\end{align}
determined by the single-particle ES $\xi(p) = \beta \omega(p)$.

\section{\label{app:EHT_QVL_summary} Entanglement Hamiltonian learning}
In the main text, we employ different protocols to find the EH, which we briefly summarize in this appendix.

\subsection{Entanglement Hamiltonian Tomography}
For the case of free fermions, we performed EHT as follows. First, we simulated projective measurements for all reduced states $\rho_{\mathbf{n}\mathbf{m}}$ supported on two neighboring lattice sites $\langle\mathbf{n}\mathbf{m}\rangle$ for a fixed number of measurements $N_\text{meas.}$ to estimate the probability $P^\text{exp.}_{{\mathbf{n}\mathbf{m}}}(U_{\mathbf{n}\mathbf{m}},s_{\mathbf{n}\mathbf{m}})$ to find a state $|s_{\mathbf{n}\mathbf{m}}\rangle$ after some unitary transformation $U_{\mathbf{n}\mathbf{m}}$ of the computational basis. For a Gaussian ansatz $\rho^\text{var.}(\boldsymbol{g})$, one can calculate the corresponding probability 
\begin{align}
P^\text{var.}_{{\mathbf{n}\mathbf{m}}}&(U_{\mathbf{n}\mathbf{m}},s_{\mathbf{n}\mathbf{m}},\boldsymbol{g}) \nonumber \\ &= \text{Tr} \left[ \rho^\text{var.}(\boldsymbol{g}) \, U_{\mathbf{n}\mathbf{m}} \, |s_{\mathbf{n}\mathbf{m}}\rangle \langle s_{\mathbf{n}\mathbf{m}}| \,  U_{\mathbf{n}\mathbf{m}}^\dagger  \right] 
\end{align}
exactly and thus find optimal parameters by minimizing
\begin{align}
    \chi^2 = \sum_{\mathbf{n}\mathbf{m}} \sum_{U_{\mathbf{n}\mathbf{m}},s_{\mathbf{n}\mathbf{m}}}  &\left[P^\text{exp.}_{\mathbf{n}\mathbf{m}}(U_{\mathbf{n}\mathbf{m}},s_{\mathbf{n}\mathbf{m}}) \right. \\ & - \left.P^\text{var.}_{\mathbf{n}\mathbf{m}}(U_{\mathbf{n}\mathbf{m}},s_{\mathbf{n}\mathbf{m}},\boldsymbol{g})\right]^2 \;, \nonumber 
\end{align}
For the results presented in the main text, we have taken a particle number basis spanned by the four states $|i_\mathbf{n},j_\mathbf{m}\rangle$ with $i,j\in\{0,1\}$ as the computational basis, which is directly accessible through quantum gas microscopy. We implemented EHT for measurements in this basis ($U^{(0)}_{\mathbf{n}\mathbf{m}}=\mathbf{1}$) and for two additional rotations
\begin{align}
    U^{(1)}_{\mathbf{n}\mathbf{m}} = e^{i\frac{\pi}{{4}} \left(c^\dagger_\mathbf{n} c_\mathbf{m} + \text{h.c.}\right)} \;, &&
    U^{(2)}_{\mathbf{n}\mathbf{m}} = e^{\frac{\pi}{{4}} \left(c^\dagger_\mathbf{n} c_\mathbf{m} - \text{h.c.}\right)} \;,
\end{align}
which can be realized by controlling the single-particle tunneling of nearest neighbours. As an alternative to EHT we employed an entropic approximation obtained by minimizing the relative entropy $S(\rho^\text{exp.}|\rho^\text{var.}(\boldsymbol{g}))$.

\subsection{Quantum Variational Learning of the EH}
For the interacting chain, we employed QVL to find the EH~\cite{kokail2021quantum}. To determine the optimal parameters, we minimized the cost function
\begin{align}
    C (\boldsymbol{\beta}) = \sum_{t_j,O} \left(\langle O\rangle_{t_j} - \langle O\rangle_0\right)^2 \;
\end{align}
for three times $t_{0/1/2}$ chosen randomly in the interval $[0,10]$ and a fourth fixed time $t_3 = 10$ (here time is measured in units of $1/[J^2 + (J')^2]=1$). The observables $\langle O \rangle_t$ are obtained by time evolution with the ansatz of Eq.~\eqref{eq:chain_ansatz}. The explicit calculations were performed for the spin representation of the model where we monitored correlations $\langle\sigma_j^x \sigma_k^x\rangle_t$ on all nearest neighbors $(j,k)$. A similar optimization is also possible for fermionic models (see, e.g., the example of a fermionic Hubbard ladder presented in \cite{kokail2021quantum}). For the simulations discussed in the main text we have neglected statistical errors from a finite number of measurements.

As discussed in \cite{kokail2021quantum}, QVL only provides the EH up to an overall scale and symmetries. This follows from the fact if $C(\boldsymbol{\beta}) = 0$ for some $H_A^\textbf{def.}(\boldsymbol{\beta})$, then the same is true for the cost function corresponding to $\beta' H_A^\text{def.}(\boldsymbol{\beta}) + Q$, where $Q$ is any symmetry of the reduced state $\rho_A$, i.e. $\left[\rho_A,Q\right]=0$. For the data presented in the main text, we have fixed the overall scale $\beta'$ by minimizing the fidelity w.r.t. to $\rho_A$. To probe the LH conjecture, one can also leave this prefactor open as it drops out when analyzing the ES as universal ratios. Given the universality of the BW-type linear deformation of the EH close to the entanglement cut, the scale factor can also be extrapolated from smaller system sizes, where another method like EHT can be performed efficiently.

\section{\label{app:12by8_results}EHT for a $12\times8$ lattice}
In this appendix, we present additional results for the model discussed in Sec.~\ref{sec:LH_IQH_CA}. We repeat the EHT analysis to demonstrate that the method is not limited to the small bulk (sub)system sizes discussed in the main text. To simplify the numerical simulations, we do not investigate the effect a finite measurement budget here and directly simulate the $N_\text{meas.} = \infty$ outcome.

\begin{figure}[t]
    \centering
    \includegraphics[width=0.9\columnwidth]{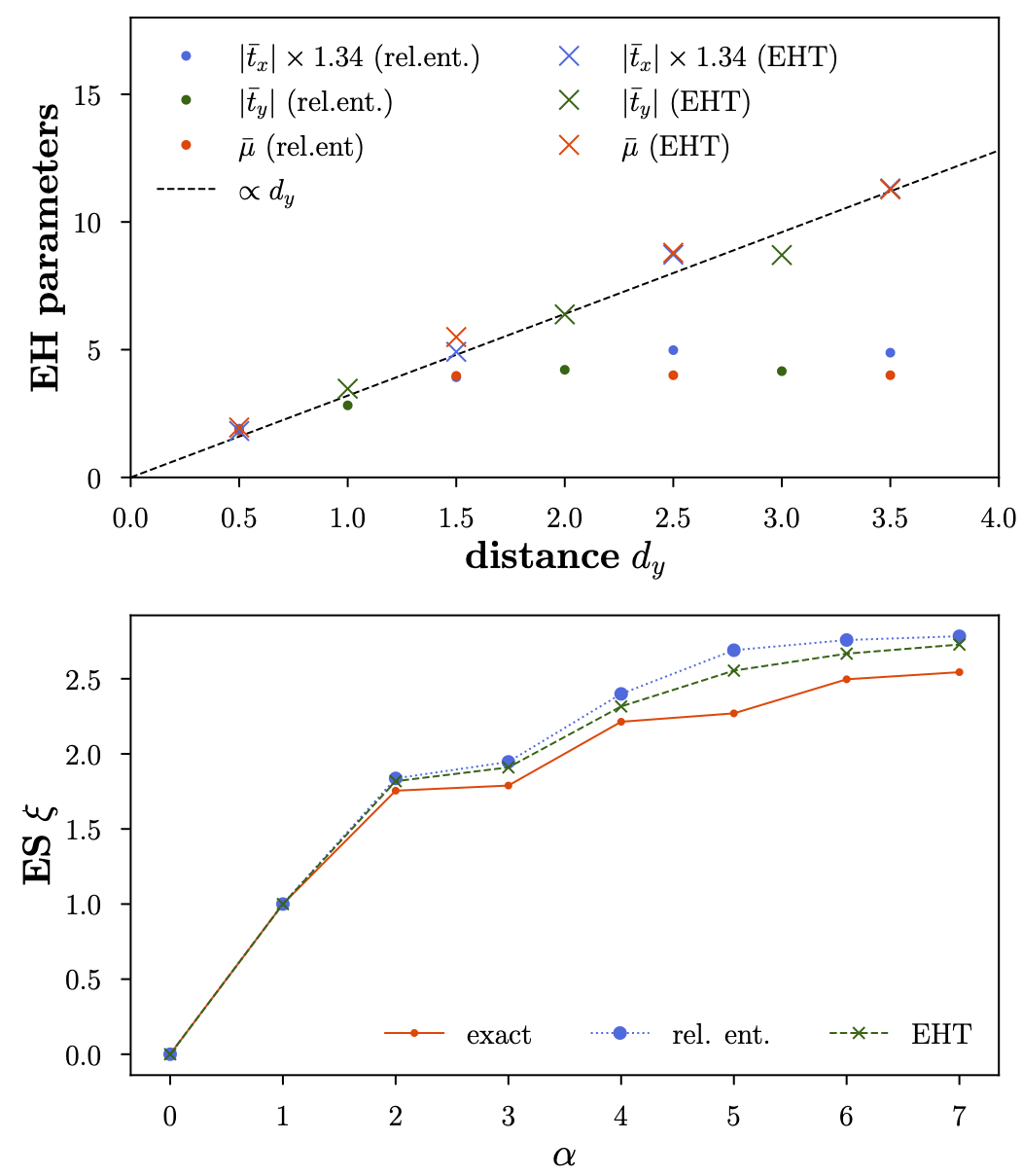}
    \caption{{EHT on a $N_x \times N_y = 12\times8$ lattice (subsystem $12\times 4$) with the same model parameters as discussed in section~\ref{subsec:IQH_EHT}. Top: The learned parameters of the EH, averaged over the $x$-direction and plotted against the distance $d_y$ from the entanglement cut. Crosses correspond to EHT (without simulated shot noise); dots are results from minimizing the relative entropy; the dashed line indicates a linear BW-type deformation. Note the rescaling of the $t_x$ parameters, such that $t_x/t_y$ matches the ratio of the original Hamiltonian. Bottom: The lowest levels of the ES corresponding to the learned deformations in comparison to the exact values.}}
    \label{fig:12by8_results}
\end{figure}

The model and its parameters are the same as discussed in section~\ref{subsec:IQH_EHT} except for the lattice geometry, which we now take as a wider ``ladder'' of size $N_x\times N_y = 12\times 8$. We study EHT for a subsystem of size $N_x^A \times N_y^A = 12\times 4$ by cutting the short direction in half.
For this geometry, we obtain a fidelity of $\mathcal{F}_Q\approx 0.95$ w.r.t.~the exact reduced state, which is again higher than the optimal result obtained from minimizing the relative entropy with $\mathcal{F}_Q\approx 0.74$. Nevertheless, both learned EHs reproduce the low-lying exact ES with comparable accuracy (see Fig.~\ref{fig:12by8_results}). Here, the onset of the typical edge mode counting $1,1,2, \dots$ is barely visible because of the relatively short $x$-direction (cf. Figs.~\ref{fig:overview}d and \ref{fig:LH_}).

We again find that the learned EH parameters take a BW-type shape. The phases of the hopping terms $t_{x/y}$ (not shown) are in excellent agreement with those of $H_\text{2D}$ [Eq.~\eqref{eq:H2D}], as already observed in Fig.~\ref{fig:optimal_pars}. With the larger $y$-direction studied in this appendix, the linear BW-type deformation of the learned parameters now also becomes more clearly visible. This is demonstrated in Fig.~\ref{fig:12by8_results}, where we plot the absolute values of the learned parameters as a function of the distance $d_y$ from the entanglement cut, averaged over the $x$-direction. Here, we define $d_y=1, 2, 3$ for the $t_y$ terms and $d_y=0.5,1.5,2.5,3.5$ for the $t_x$ and $\mu$ terms. We find that both entropic and EHT parameters exhibit a linear slope close to the entanglement cut, in remarkable agreement with our expectation from the BW theorem. While the former quickly bends off around $d_y=1.5$ to an approximately constant value, the EHT deformation approximately follows the linear slope throughout the whole subsystem.

\section{\label{app:finite_temp}The effect of finite temperature}
In this appendix, we present another set of additional results for the model discussed in Sec.~\ref{sec:LH_IQH_CA}. Here we study the effect of a finite temperature $T = 1/\beta$.
Our main goal is to demonstrate that the quasi-local ansatz remains applicable for $\beta < \infty$ and to investigate the impact of finite temperature on the reconstructed ES. We note that we keep using the names ``Entanglement'' spectrum and ``Entanglement'' Hamiltonian for the learned Hamiltonian and its spectrum, even though they lose their direct interpretation in terms of entanglement at finite temperature (or for any other mixed state).

We study again the free fermion model on a $N_x \times N_y$ ladder with the same parameters as in section~\ref{subsec:IQH_EHT}. We consider a grand-canonical ensemble, i.e. the globally mixed state
\begin{align}
    \rho(\beta, \mu) \propto e^{-\beta H_\text{2D} (\mu)} \;,
\end{align}
with the prefactor fixed by normalization $\text{Tr} \left[\rho(\beta,\mu)\right] = 1$. In order to compare results at different finite temperatures, we tune the chemical potential $\mu = \mu(\beta)$ in $H_\text{2D}$ [Eq.~\eqref{eq:H2D}] such that the average of the total fermion number operator $N_F$ is fixed at filling $1/3$,
\begin{align}
    \text{Tr} \left[\rho(\beta,\mu) N_F \right] \overset{!}{=} N_x N_y/3 \;.
\end{align}
In this way, we obtain a familiy of states $\rho(\beta) \equiv \rho(\beta, \mu(\beta))$ that converges to the ground state studied in the main text in the limit $\beta \rightarrow\infty$. For these states, we consider the reduced density matrices $\rho_A (\beta) \equiv e^{-\tilde{H}_A(\beta)}$ on the same subsystems $A$ as discussed in the main text, which defines a finite-temperature analog of the ``Entanglement'' Hamiltonian, $\tilde{H}_A(\beta)$. With this setup, we repeat our EH learning analysis. 
For simplicity, we restrict ourselves to the entropic method.

Throughout this article, we have seen that at zero temperature the EH can be well approximated by a BW-type deformation. Since the bulk of the system considered here is in a gapped state with energy gap $\Delta \sim |t_{x/y}|$, we expect that this behaviour will remain qualitatively unchanged as long as the temperature $T=1/\beta$ is sufficiently low, i.e. $T \ll \Delta \sim \mathcal{O}(1)$. At high temperatures, $T\gg \Delta$, we expect that typical subsystems of the thermal state are again well approximated by a thermal state with the same temperature. Since also this limit is described by a quasi-local EH, we expect that our ansatz remains applicable at all temperatures~\footnote{{Here we disregard potential phase transitions.}}.

\begin{figure}[t]
	\centering
	\includegraphics[width=\columnwidth]{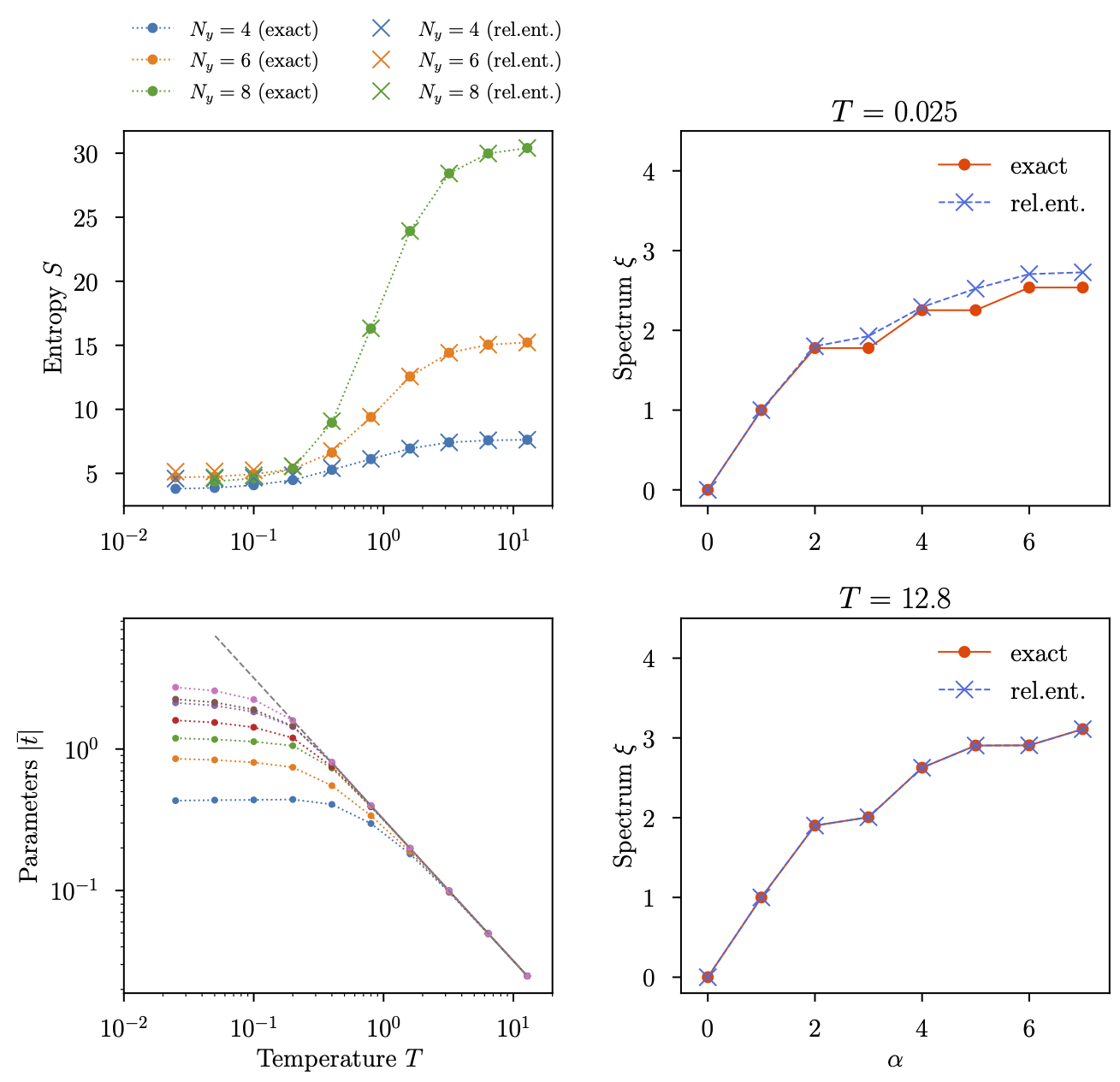}
	\caption{{Learning the ``Entanglement'' Hamiltonian at finite temperature. Top left: The exact values of the von Neumann entropy (circles) of the reduced state $\rho_A$ are perfectly  reproduced by the optimal quasi-local approximation (crosses) of the EH. Bottom left: The absolute values of the learned hopping amplitudes $t$ as a function of temperature. Different colors represent different distances from the entanglement cut (small values are closer to the cut), averaged over the $x$-direction of the ladder (see also appendix~\ref{app:12by8_results}), and the dashed grey line indicates the expected $1/T$ behaviour at large temperature. Right: spectrum of the learned EH at low (top) and high (bottom) temperature in comparison to the exact results. }}
	\label{fig:finite_temp}
\end{figure}

These expectations are confirmend by our numerical simulations, summarized in Fig.~\ref{fig:finite_temp}. As an overall check of the quality of the approximation, we find that the von Neumann entropy of the reduced state is precisely reproduced by the learned EH. Here, we have studied several ladder widths $N_y=4,6,8$, which indicates that the quasi-local ansatz is capable of interpolating between the low-temperature area law (i.e. approximately constant entropy $S \sim \text{const}$ for fixed length $N_x^A = 12$) to the high-temperature volume law (i.e. approximately linear dependence $S \propto N_y^A$ with the subsystem width $N_y^A$). Furthermore, we find that the EH takes a BW-type deformation for all temperatures. This is illustrated in Fig.~\ref{fig:finite_temp} for the tunneling terms, whose absolute values are shown for all available distances from the entanglement cut for the largest subsystem of width $N_y^A = 4$. The expected approximately linear slope of the $T=0$ limit flattens with increasing temperature until the values collapse around $T\sim 1$, consistent with the expectation of an ordinary thermal state. Finally, Fig.~\ref{fig:finite_temp} also shows the ES for two examples of a large and small temperature, again for the largest subsystem studied. In general, we find that the agreement of the low-lying ES is much better at high temperatures where the reduced state becomes a simple thermal state. 

We also note that finite-temperature fluctuations in general break the clear degeneracy pattern $1,1,2, \dots$ of the ES. However, for the present model this effect is barely visible in Fig.~\ref{fig:finite_temp}, where relatively small system sizes are shown. Moreover, the non-rescaled ES has much smaller spacings at higher temperatures because the state becomes increasingly mixed, approaching the maximal mixed state as $T\rightarrow \infty$, which is also reflected in the increasing entropy. As a consequence, there exists a finite temperature beyond which it becomes practically impossible to resolve the degeneracy pattern in an experimental setup with finite precision (e.g. due to a finite measurement budget).

In summary, we have found that the quasi-local ansatz underlying our proposal is not restricted to zero temperature and instead remains applicable at finite $T>0$, which indicates that also our proposal to test the LH conjecture remains applicable. As a final note of caution, we want to emphasize that strictly speaking the EH and ES then lose their ``entanglement meaning'' since, e.g., the von Neuman entropy also contains a purely classical contribution from thermal fluctuations. In this context, our proposal might be extended to the recently introduced ``Negativity Hamiltonian''~\cite{murciano2022negativity}. For future experimental realization of our proposal -- where some amount of finite temperature fluctuations are practically unavoidable -- the results presented in this appendix indicate that the zero temperature behaviour can be faithfully probed at sufficiently low temperatures $T \lesssim \Delta$, set by the mass gap $\Delta$ of the insulating bulk.

\bibliographystyle{quantum}
\bibliography{references}

\end{document}